\newif\ifExtended
\Extendedtrue
\newcommand{\ifextension}[2]{\ifExtended%
#1%
\else%
#2%
\fi}

\ifextension{
\documentclass[conference]{IEEEtran}
}{
\documentclass[year=25,pdfa]{fmcad}
}
\IEEEoverridecommandlockouts
\usepackage{cite}
\usepackage{amsmath,amssymb,amsfonts}
\usepackage{algorithmic}
\usepackage{graphicx}
\usepackage{textcomp}
\usepackage{xcolor}
\usepackage{paralist}
\usepackage{hyperref}
\usepackage{cleveref}
\crefname{section}{Sec.}{Secs.}
\usepackage[T1]{fontenc}
\usepackage{graphicx}

\usepackage{graphicx}
\usepackage{tikz}
\usetikzlibrary{arrows,decorations.pathreplacing, positioning}
\usepackage{pgfplots, pgfplotstable}
\usetikzlibrary{shapes.geometric, arrows, fit, positioning}
\usetikzlibrary{arrows.meta,patterns}

\usetikzlibrary{decorations.pathmorphing}
\usetikzlibrary{fpu}
\usetikzlibrary{tikzmark}
\usetikzlibrary{calc}
\pgfplotsset{compat=1.18}
\usepgfplotslibrary{fillbetween}
\tikzset{
  pics/robot/.style args={}{
  code={
    \begin{scope}[
      every node/.style={
        draw,
        font=\footnotesize,
        node distance=3mm and 4mm
      }
    ] 
            \draw[very thick, rounded corners=0.5ex,fill=white,thick]  (0,.15) -- ++(2,0) -- ++(0,1.0) -- ++(-0.5,0) -- ++(0.0,0.75) -- ++(-1.0,0.0) -- ++(0,-0.75) -- ++(-0.5, 0.) -- ++(0,-1.0);
            \draw[draw=black,fill=black,thick] (0.75,1.6) circle (.05);
            \draw[draw=black,fill=black,thick] (1.25,1.6) circle (.05);
            \draw[decorate,decoration={zigzag,segment length=1mm,amplitude=0.3mm}] (0.75,1.3) -- (1.25,1.3);

            \draw[decorate,decoration={zigzag,segment length=1mm,amplitude=0.3mm}] (0.75,1.9) -- (0.65,2.4);
            \draw[draw=black,fill=black,thick] (0.65,2.4) circle (.05);
            \draw[decorate,decoration={zigzag,segment length=1mm,amplitude=0.3mm}] (1.25,1.9) -- (1.35,2.4);
            \draw[draw=black,fill=black,thick] (1.35,2.4) circle (.05);
            
            \draw[draw=black,fill=gray!50,thick] (0.4,.165) circle (.33);
            \draw[draw=black,fill=gray!50,thick] (1.5,.165) circle (.33);
            \draw[draw=black,fill=black,thick] (0.4,.165) circle (.05);
            \draw[draw=black,fill=black,thick] (1.5,.165) circle (.05);
        \end{scope}
    }
  }
}
\tikzset{
  pics/car/.style args={}{
  code={
    \begin{scope}[
      every node/.style={
        draw,
        font=\footnotesize,
        node distance=3mm and 4mm
      }
    ] 
            \draw[very thick, rounded corners=0.5ex,thick]  (0,.2) -- ++(2,0) -- ++(0,0.3) -- ++(-0.5,0.1) -- ++(-0.25,+0.25) -- ++(-0.75,0.0) -- ++(-0.25,-0.15) -- ++(-0.25, 0.) -- (0,.2);
            \draw[draw=black,fill=gray!50,thick] (0.4,.2) circle (.2);
            \draw[draw=black,fill=gray!50,thick] (1.5,.2) circle (.2);
            \draw[draw=black,fill=black,thick] (0.4,.2) circle (.035);
            \draw[draw=black,fill=black,thick] (1.5,.2) circle (.035);
        \end{scope}
    }
  }
}\tikzstyle{nnnode}=[thick,circle,minimum size=0.3cm,inner sep=0.3,outer sep=0.6,draw=black,fill=gray]
\tikzstyle{nnnode in}=[nnnode,draw=blue,fill=blue]
\tikzstyle{nnnode hidden}=[nnnode,draw=orange,fill=orange]
\tikzstyle{nnnode out}=[nnnode,draw=cyan,fill=cyan]
\tikzset{ %
nnnode 1/.style={nnnode in},
nnnode 2/.style={nnnode hidden},
nnnode 3/.style={nnnode hidden},
nnnode 4/.style={nnnode out}
}
\usepackage{listofitems}

\usepackage{amsmath}
\usepackage{amsfonts}
\usepackage{mathtools}
\usepackage{stmaryrd}
\usepackage{mleftright}
\usepackage{wrapfig}
\usepackage{amsthm}
\newtheorem{definition}{Definition}

\newtheorem{lemma}{Lemma}

\usepackage[inline]{enumitem}

\usepackage{hyperref}

\usepackage{cleveref}

\usepackage{multirow}

\crefname{formula}{formula}{formulas}
\creflabelformat{formula}{#2(#1)#3}

\crefname{lemma}{lemma}{lemmas}

\usepackage[createShortEnv]{proof-at-the-end}
\pgfkeys{/prAtEnd/global custom defaults/.style={
one big link={See proof in Appendix.}
}
}

\newcommand{\reals}{\ensuremath{\mathbb{R}}}
\newcommand{\robotPos}{\ensuremath{p}}
\newcommand{\robotVel}{\ensuremath{v}}

\newcommand{\stateOne}{\ensuremath{\nu}}
\newcommand{\stateTwo}{\ensuremath{\mu}}
\newcommand{\stateSpace}{\mathcal{S}}

\newcommand{\dL}{\ensuremath{d\mathcal{L}}}
\newcommand{\dGL}{\ensuremath{dG\mathcal{L}}}

\renewcommand{\implies}{\rightarrow}
\renewcommand{\iff}{\leftrightarrow}

\newcommand{\precondition}{\ensuremath{\mathrm{pre}}}
\newcommand{\postcondition}{\ensuremath{\mathrm{post}}}
\newcommand{\environmentProgram}{\ensuremath{\mathrm{env}}}
\newcommand{\controllerProgram}{\ensuremath{\mathrm{ctl}}}
\newcommand{\preProgram}{\ensuremath{\mathrm{angel}_{\text{pre}}}}
\newcommand{\preProgramParam}[1]{\ensuremath{\mathrm{angel}_{\text{pre}}^{#1}}}
\newcommand{\postProgram}{\ensuremath{\mathrm{angel}_{\text{post}}}}
\newcommand{\postProgramParam}[1]{\ensuremath{\mathrm{angel}_{\text{post}}^{#1}}}
\newcommand{\envelopeExactOneDirOneWall}{\ensuremath{\controllerProgram_1^1}}
\newcommand{\envelopeExactTwoDirOneWall}{\ensuremath{\controllerProgram_1^2}}
\newcommand{\envelopeExactTwoDirTwoWall}{\ensuremath{\controllerProgram_2^2}}
\newcommand{\postExactOneWall}{\ensuremath{\postcondition_1}}
\newcommand{\postExactTwoWall}{\ensuremath{\postcondition_2}}
\newcommand{\preExactOneWall}{\ensuremath{\precondition_1}}
\newcommand{\preExactTwoWall}{\ensuremath{\precondition_2}}

\newcommand{\robotSimplePerturb}{\ensuremath{\alpha_{\text{noise}}}}

\newcommand{\safetyPredicateSym}{\ensuremath{\mathrm{safety}}}
\newcommand{\safetyPredicate}[1]{%
\ensuremath{%
\safetyPredicateSym\mleft(%
#1
\mright)%
}%
}%

\newcommand{\controllerMonitor}{\ensuremath{\chi_{\mathrm{ctrl}}}}
\newcommand{\monitorExactOneDirOneWall}{\controllerMonitor^{1,1}}

\newcommand{\monitorExactTwoDirTwoWall}{\controllerMonitor^{2,2}}

\newcommand{\implementationPredicateSym}{\ensuremath{\mathrm{impl}}}
\newcommand{\implementationPredicate}[2]{\ensuremath{\implementationPredicateSym{}\mleft(#1,#2\mright)}}
\newcommand{\implementationPredicateRobotExactSym}{\implementationPredicateSym{}_{\mathbb{R}}}
\newcommand{\implementationPredicateRobotExact}[2]{\ensuremath{\implementationPredicateRobotExactSym{}\mleft(#1,#2\mright)}}

\newcommand{\stateFormula}{\ensuremath{\chi_{\mathrm{inv}}}}
\newcommand{\stateExactOneDirOneWall}{\stateFormula^{1,1}}
\newcommand{\stateExactTwoDirTwoWall}{\stateFormula^{2,2}}

\newcommand{\preMonitor}[0]{\ensuremath{\chi_{\mathrm{pre}}}}
\newcommand{\postMonitor}[0]{\ensuremath{\chi_{\mathrm{post}}}}

\def\BibTeX{{\rm B\kern-.05em{\sc i\kern-.025em b}\kern-.08em
    T\kern-.1667em\lower.7ex\hbox{E}\kern-.125emX}}
\begin{document}

\title{Of Good Demons and Bad Angels:\\
Guaranteeing Safe Control under Finite Precision
\thanks{This work was supported by funding from the pilot program Core-Informatics of the Helmholtz Association (HGF).}
}

\ifextension{
\author{
\IEEEauthorblockN{Samuel Teuber, Debasmita Lohar, and Bernhard Beckert }
   \IEEEauthorblockA{Karlsruhe Institute of Technology, Karlsruhe, Germany
   \\\{teuber, debasmita.lohar, beckert\}@kit.edu}
   }
}{
\author{\IEEEauthorblockN{1\textsuperscript{st} Samuel Teuber \orcid{0000-0001-7945-9110}}
\IEEEauthorblockA{\textit
\textit{Karlsruhe Institute of Technology}\\
Karlsruhe, Germany \\
teuber@kit.edu}
\and
\IEEEauthorblockN{2\textsuperscript{nd} Debasmita Lohar \orcid{0000-0001-8639-4116}}
\IEEEauthorblockA{
\textit{Karlsruhe Institute of Technology}\\
Karlsruhe, Germany \\
debasmita.lohar@kit.edu}
\and
\IEEEauthorblockN{3\textsuperscript{rd} Bernhard Beckert \orcid{0000-0002-9672-3291}}
\IEEEauthorblockA{
\textit{Karlsruhe Institute of Technology}\\
Karlsruhe, Germany \\
beckert@kit.edu}
}
}
\maketitle

\begin{abstract}
As neural networks (NNs) become increasingly prevalent in safety-critical neural network-controlled cyber-physical systems (NNCSs), formally guaranteeing their safety becomes crucial. 
For these systems, safety must be ensured throughout their entire operation, necessitating infinite-time horizon verification. To verify the infinite-time horizon safety of NNCSs, recent approaches leverage Differential Dynamic Logic (\dL{}). However, these \dL{}-based guarantees rely on idealized, real-valued NN semantics and fail to account for roundoff errors introduced by finite-precision implementations.

\looseness=-1
This paper bridges the gap between theoretical guarantees and real-world implementations by incorporating robustness under finite-precision perturbations—in sensing, actuation, and computation—into the safety verification. We model the problem as a \emph{hybrid game} between a \emph{good Demon}, responsible for control actions, and a \emph{bad Angel}, introducing perturbations. This formulation enables formal proofs of robustness w.r.t. a given (bounded) perturbation. Leveraging this bound, we employ state-of-the-art mixed-precision fixed-point tuners to synthesize sound and efficient implementations, thus providing a complete end-to-end solution. We evaluate our approach on case studies from the automotive and aeronautics domains, producing efficient NN implementations with rigorous infinite-time horizon safety guarantees.
\end{abstract}


\section{Introduction}
\label{sec:intro}
\looseness=-1
Neural networks (NNs) are increasingly integrated into critical systems, such as adaptive cruise control in cars~\cite{DBLP:conf/aaai/FultonP18,Brosowsky2021,ARCH21:ARCH_COMP21_Category_Report_Artificial} or collision avoidance in airplanes~\cite{JulianKDVCASSafe,julian2016policy}. Thus, ensuring the safety of NN Control Systems (NNCSs) becomes imperative.

Recent research has made significant progress toward this goal. Some approaches~\cite{zhang2018efficient,xu2020automatic,xu2020fast,wang2021beta,Henriksen2020,Henriksen2021,katz2017reluplex,katz2019marabou,bunel2020branch,bak2020improved,bakOverapprox,zhang2022general,shi2024genbab,DBLP:conf/cav/WuIZTDKRAJBHLWZKKB24} only analyze the input-output behavior of the NN but ignore the essential feedback-loop dynamics between the NN and its physical environment. Others~\cite{Forets2019JuliaReachReachability,Schilling2021,tran2019star,tran2020neural,9093970,IvanovACM21,Ivanov2021,Huang2019,Fan2020,Dutta2019,Akintunde2022,Sidrane2021} consider this feedback loop but only verify safety for a \emph{finite}-time horizon, which may be insufficient.
A recent technique, VerSAILLE~\cite{Teuber2024}, addresses this limitation by proving safety throughout the entire operation (\emph{infinite-time horizon}) of the NNCS using Differential Dynamic Logic (\dL{}).

However, VerSAILLE's strong guarantees rely on idealized, real-valued NN semantics and do not account for roundoff errors arising from practical finite-precision (floating- or fixed-point) implementations. Finite-precision arithmetic introduces errors at potentially every operation, which can accumulate and lead to incorrect or unsafe decisions~\cite{lohar2018discrete}. Furthermore, efficiency demands often require implementations in low-precision fixed-point arithmetic~\cite{gopinath2019compiling,gupta2015deep}, introducing larger numerical errors in the computation. Thus, to ensure real-world safety, it is crucial to verify that safety guarantees are valid under finite-precision semantics.

While verification of quantized NNs has been explored~\cite{henzinger2021scalable}, it is known to be PSPACE-hard and typically focuses on input-output specifications rather than closed-loop dynamics, which must simultaneously account for the continuous, real-valued behavior of physical dynamics.
Sound quantization techniques~\cite{Aster} automate precision tuning but leave the burden of safety verification to the user. A recent work~\cite{harikishan2024towards} has combined safety analysis with bounded roundoff errors, but their guarantees are limited to finite-time horizons.

This leaves a critical gap: the need for an efficient and practical method to guarantee \emph{infinite-time horizon safety} of NNCSs while accounting for finite-precision errors.

In this paper, we propose a general infinite-horizon safety verification methodology, building on VerSAILLE~\cite{Teuber2024}, that accounts for finite-precision errors in sensing, actuation, and computation. 
We observe that some "safe" \dL{} control strategies (\textit{envelopes}) can become unsafe even under miniscule input or output perturbations, making them unsuitable for verifying realistic (NN) controllers. To address this, we introduce a new notion of \emph{robustness} for control envelopes, which is also of independent interest beyond handling roundoff errors. Finally, we integrate our approach with mixed-precision tuning to generate sound and efficient fixed-point implementations that can be directly compiled using standard high-level synthesis tools for FPGAs, providing a complete end-to-end solution. To summarize, we make the following contributions:

\begin{enumerate}[label=(C\arabic*)]
\item We formalize robustness for control envelopes under perturbation -- a necessary condition for finite-precision \dL{}-based NN verification -- using Differential Game Logic (\dGL{}), %
as a hybrid game between a \emph{good} Demon (control actions) and a \emph{bad} Angel (perturbations).
We present a decidable, sufficient criterion for robustness.

\item
We propose a novel, decidable criterion
for guaranteeing the infinite-time horizon safety of NNs 
under given
perturbations leveraging
existing \emph{real-valued} NN verifiers.

\item We integrate precision tuning (with Daisy~\cite{Daisy}) to synthesize efficient mixed-precision fixed-point implementations that guarantee a provided error bound and are compatible with standard hardware synthesis tools~\cite{websiteOfVivado}. 
\end{enumerate}
We evaluated our approach on three realistic case studies: an Adaptive Cruise Control (ACC) system~\cite{DBLP:conf/aaai/FultonP18,Teuber2024} with discrete and continuous acceleration control, and a Vertical Airborne Collision Avoidance System (VCAS)\cite{JulianKDVCASSafe}. Our analysis shows that the continuous ACC control envelope from\cite{Teuber2024} is not robust to output perturbations; we propose modifications and verify safety under bounded errors. For VCAS, our results imply (limited) robustness of the original control envelope. Using the computed perturbation bounds, we synthesized mixed-precision fixed-point implementations for all case studies, compiled them with Xilinx Vivado~\cite{websiteOfVivado}, and measured their latency (running time) in machine cycles.

\begin{figure*}[t]
    \centering
    \begin{tikzpicture}
    \node (model) [anchor=north west,align=center,draw=black] at (0,0) {\textbf{$\mathbf\dL$ Model}\\{\color{green!50!black} safe}};

    \node (perturbation) [anchor=north west,align=center,draw=black] at (2.25,-1.5) {%
    \textbf{Angelic Perturbations}\\
    modeled as hybrid programs
    };

    \node (nnspec) [anchor=north west,align=center,draw=black] at (0, -4) {
    \textbf{NN Specification}
    };

    \node (robustness) [anchor=north west,align=center,draw=black,minimum width=10.75cm] at (7.25,0) {
    \textbf{Robustness of Control Envelopes (\Cref{thm:robust_envelope})}\\
    \makebox[9cm][c]{
    \begin{tikzpicture}[node distance=0]
        \node (demon) {
        \includegraphics[width=0.75cm]{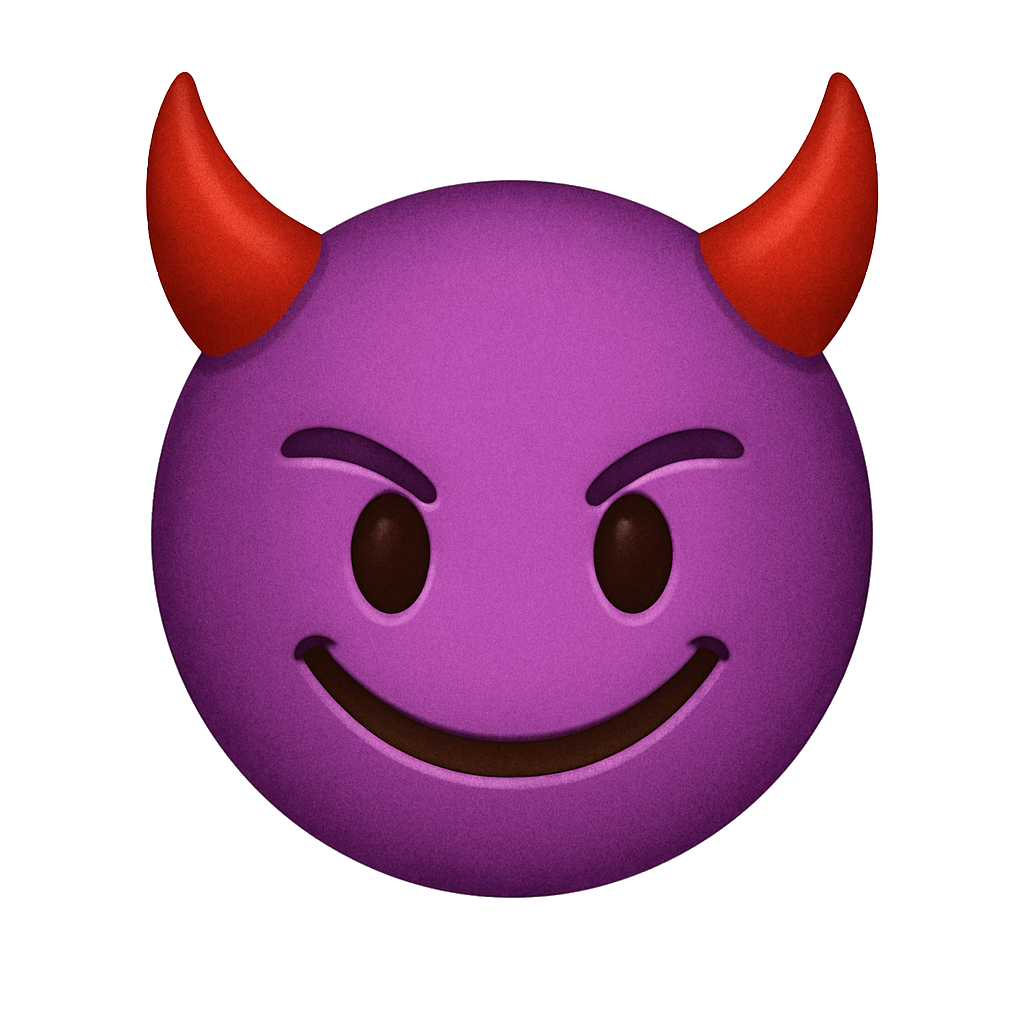}
        };
        \node [below=of demon,align=center] {\textbf{good demon}\\chooses control decisions};
        \node (vs) [right of=demon,node distance=2cm] {\textbf{vs.}};
        \node (angel) [right of=vs,node distance=2cm] {
        \includegraphics[width=0.75cm]{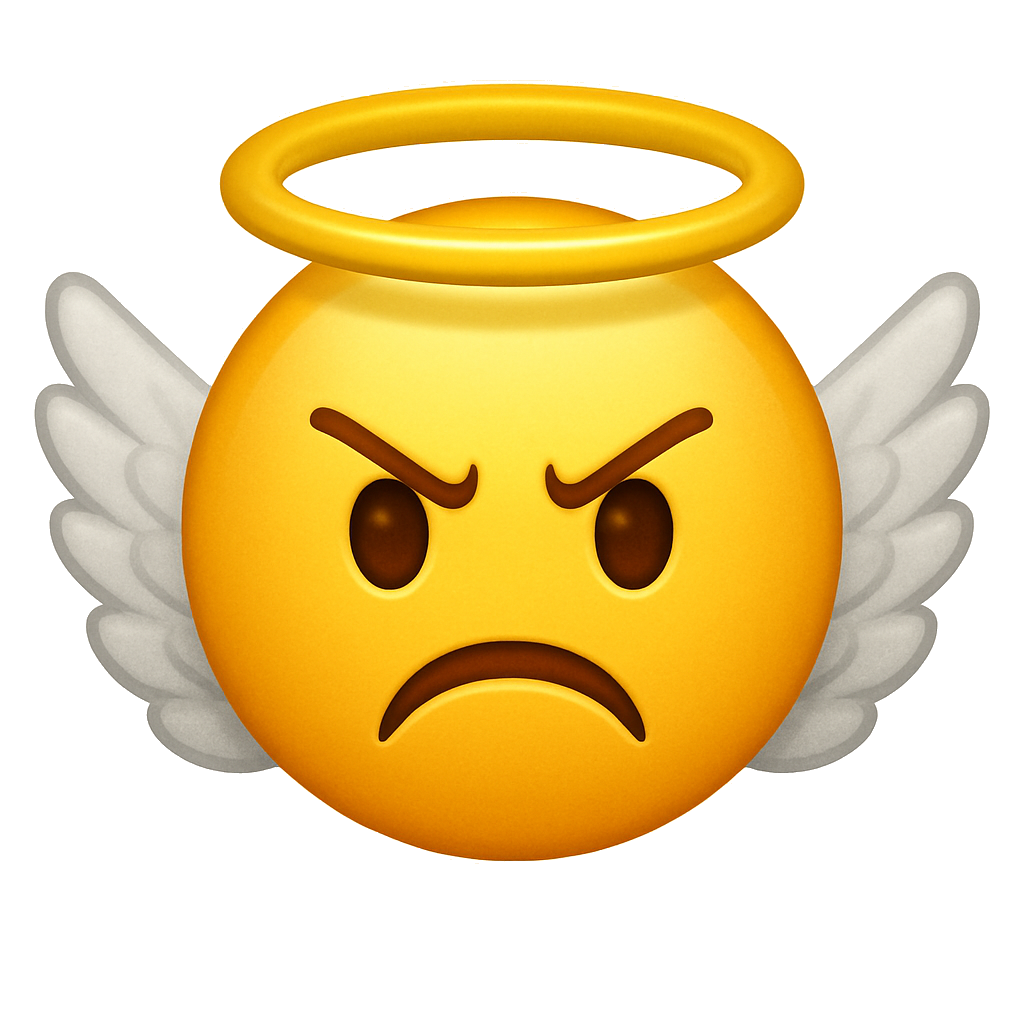}
        };
        \node [below=of angel,align=center] {\textbf{bad angel}\\
        chooses perturbations};
    \end{tikzpicture}}\\
    robustness $\equiv$ ``demon has winning strategy''\\[0.25cm]
    {sufficient conditions checkable via real arithmetic SMT}
    };

    \node (perturbationSafe) [anchor=north west,align=left,draw=black,minimum width=10.5cm,text width=10.5cm,minimum height=1.75cm] at (7.25,-4) {
    \begin{minipage}{7cm}
    \begin{center}
    \textbf{Safety under Perturbation (\Cref{thm:correctness_under_perturbation})}\\[0.2cm]
    $\equiv$\\
    ``implements winning strategy for demon''
    \end{center}
    \end{minipage}
    };

    \node (synthesis) [anchor=north west,align=left,draw=black,minimum width=10.5cm,text width=10.5cm,minimum height=1.75cm] at (7.25,-5.75) {
    \begin{minipage}{7cm}
    \begin{center}
    \textbf{Sound Quantization (\Cref{subsec:method:implementation})}\\[0.2cm]
    synthesis of fixed-point NN implementation
    \end{center}
    \end{minipage}
    };

    \draw [->] (model.south) -- node[align=left,rotate=90,anchor=south] {\footnotesize invariant \&\\[-0.35em]
    \footnotesize controller monitor\\[-0.35em]
    \footnotesize (VerSAILLE~%
    {\protect\NoHyper\cite{Teuber2024}\protect\endNoHyper})} (model.south|-nnspec.north);
    \draw [->] (perturbation.west) -| node [pos=0.75,anchor=south,rotate=90] {\footnotesize ModelPlex~%
    {\protect\NoHyper\cite{DBLP:journals/fmsd/MitschP16}\protect\endNoHyper}
    } ([xshift=0.5cm]nnspec.north);

    \draw [->] (model.east) -- (robustness.west|-model.east);
    \draw [->] (perturbation.north) |- ([yshift=-0.5cm]robustness.west|-model.east);
    \draw [->] (nnspec) -- (perturbationSafe.west|-nnspec);
    \node [anchor=north west,text width=6cm,align=left] at(0,-5) {
    \footnotesize
    We analyze a given $\dL$ model w.r.t. perturbations modelled as \emph{angelic perturbations} and use this perturbation model in three ways:
    (i)~to check whether a provably safe \dL{} control envelope is robust to perturbations;
    (ii)~to verify that the NN is safe under such perturbations; and
    (iii)~to synthesize a fixed-point implementation of the NN whose error stays within the modelled perturbation bounds.
    };

    \draw [<->] (perturbation.east) -- ([xshift=0.25cm]perturbation.east) |- node [pos=0.42,rotate=90,anchor=south] {\footnotesize matching error bounds} (synthesis.west);

    \node (nn) [anchor=north west,align=center,minimum width=3.75cm, minimum height=3.5cm,draw=black] at (14.25,-4) {
    \begin{tikzpicture}[x=1.5cm,y=3cm,scale=0.3,
    ]

\readlist\Nnod{1,3,3,2} %

\foreachitem \N \in \Nnod {
  \foreach \i [evaluate={
      \y=0.5*(\i - (\N*0.5));%
      \x=1.5*\Ncnt;
      \prev=int(\Ncnt-1);
  }] in {0,...,\N} {
    \node[nnnode \Ncnt, minimum size=0.2cm,anchor=center] (N\Ncnt-\i) at (\x,\y) {};

    \ifnum\Ncnt>1
      \foreach \j in {0,...,\Nnod[\prev]} {
        \path let \p1 = (N\prev-\j), \p2 = (N\Ncnt-\i),
                  \n1 = {min(\y1,\y2)},
                  \n2 = {max(\y1,\y2)},
                  \n3 = {(\n1 < 0 && \n2 > 0) ? 1 : 0},
                  \n4 = {(\n2 <= 0) ? 1 : 0},
                  \n5 = {(\y1 >= 0) ? 1 : 0},
                  \n6 = {(\y1 != \y2) ? (\y2-\y1) : 1},
                  \n7 = {(-\y1*(\x2-\x1)/(\n6)+\x1)}
        in {
        \pgfextra{\xdef\crossesZero{\n3}}
        \pgfextra{\xdef\belowZero{\n4}}
        \pgfextra{\xdef\t{1}}
        \pgfextra{\xdef\yOneAbove{\n5}}
        \pgfextra{\xdef\midPoint{\n7}}
        };

          \ifthenelse{\equal{\crossesZero}{1pt}}{

            \ifthenelse{\equal{\yOneAbove}{1pt}}{
                \draw[thick] (N\prev-\j) -- (\midPoint,0);
                \draw[thick,dashed](\midPoint,0) -- (N\Ncnt-\i);
            }{
                
                \draw[thick,dashed] (N\prev-\j) -- (\midPoint,0);
                \draw[thick](\midPoint,0) -- (N\Ncnt-\i);
            }
            
          }{
            \ifthenelse{\equal{\belowZero}{1pt}}{
              \draw[thick,dashed] (N\prev-\j) -- (N\Ncnt-\i);
            }{
              \draw[thick] (N\prev-\j) -- (N\Ncnt-\i);
            }
          }
      }
    \fi
  }
}
\end{tikzpicture}

    };
    \node [anchor=north west,align=center,minimum width=3.75cm] at (14.25,-4) {\small{real arithmetic}\\\small verification ($N^3V$~{\protect\NoHyper\cite{Teuber2024}\protect\endNoHyper})};
    \node [anchor=south west,align=center,minimum width=3.75cm] at (14.25,-7.5) {\small {mixed fixed-point}\\\small tuning (Daisy~%
    {\protect\NoHyper\cite{Daisy}\protect\endNoHyper})};

    \end{tikzpicture}
    \caption{Overview of approach which assumes a safe $\dL$ model as foundation (emojis generated by ChatGPT)}
    \label{fig:overview}
\end{figure*}
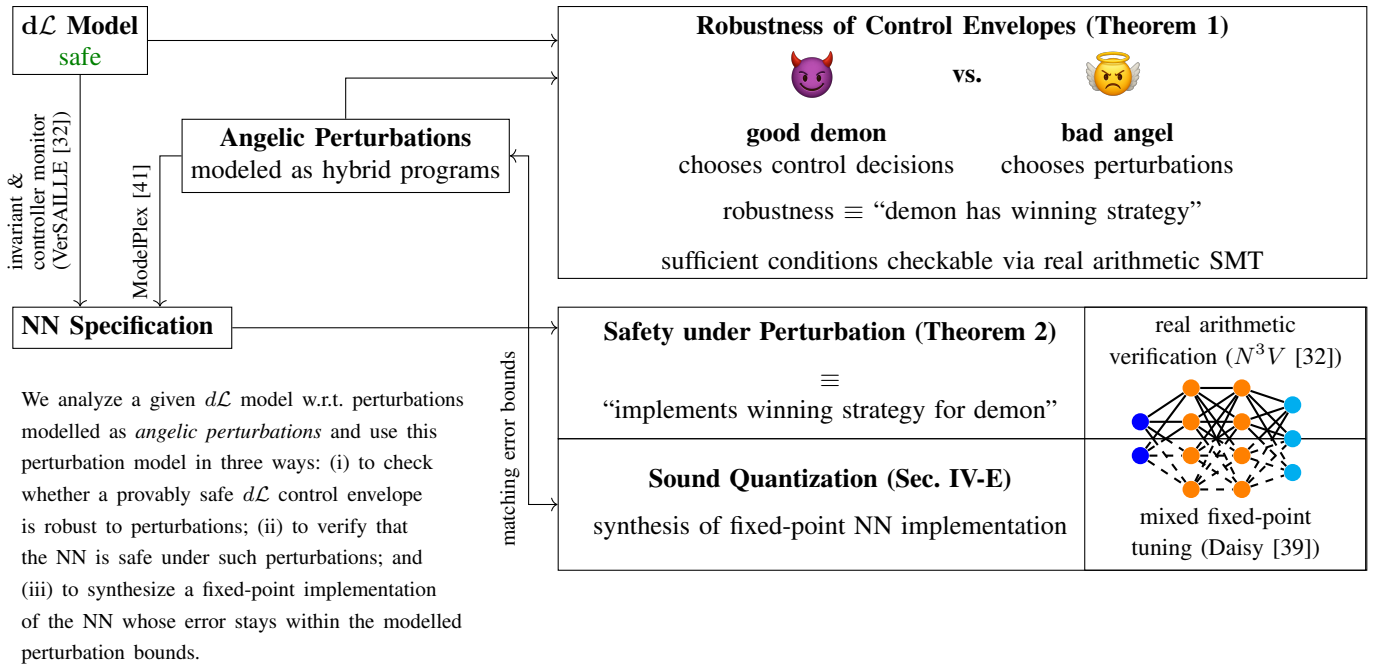
\section{Approach at a Glance}\label{overview}
This section provides a high-level overview using a simplified ground robot from~\cite{DBLP:conf/pldi/BohrerTMMP18}. The technical background and the proposed approach are presented in the subsequent sections.

\paragraph{Ground Robot Example}
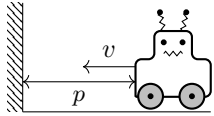
\begin{wrapfigure}[8]{r}{0.2\textwidth}
    \centering
    \begin{tikzpicture}%
    
    \pic [scale=0.5] at (1.5,0.125) {robot={}};

    \draw[-] (0,0) -- (2.5,0);
    \draw[<->] (0,0.4) -- node[anchor=north] {\small$\robotPos$} (1.5,0.4);

    \draw [->] (1.5,.6) -- node[anchor=south] {$\robotVel$} (0.8,0.6);

    \draw (0,0) -- (0,1.45);
    \fill[pattern=north west lines] (0,0) rectangle ++(-0.2,1.45);

    \end{tikzpicture}
    \caption{%
    A robot at distance $p$ controls its velocity $v$ and avoids an obstacle ($p \geq 0$).}
    \label{fig:running:robot_idea}
\end{wrapfigure}

The robot, at distance $\robotPos \in \reals$ to an obstacle, moves along a corridor (see \Cref{fig:running:robot_idea}). 
It controls velocity $\robotVel$, which can be updated once per control cycle of duration of at most $T$. To ensure safety, the robot must maintain $p \geq 0$.

Let us consider the case where the robot can only move toward the wall (i.e., $v \geq 0$).
Since the robot has full control over its velocity, a safe strategy is to choose any $v \geq 0$ such that $p - Tv \geq 0$, ensuring it does not collide with the obstacle within one control cycle. The aim is to formally prove that the robot remains safe under this strategy.

\paragraph{Nomenclature}
As we discuss control strategies at different levels, we unify the terminology as follows:
First, nondeterministic control strategies modeled and verified in \dL{} are referred to as \emph{control envelopes}.  Second, concrete real-valued NNs that deterministically provide a control action for a given input are called \emph{idealized implementations}, and will be verified against control envelopes.
Finally, hardware-level realizations of idealized implementations using fixed-point arithmetic are termed \emph{fixed-point implementations}.
We will bound their numerical errors w.r.t. their idealized counterparts.

\paragraph{Verifying Control under Finite Precision}
\label{par:glance:verifying_in_finite_precison}
A control envelope, where $v$ is chosen arbitrarily such that $p - Tv \geq 0 \land v \geq 0$, %
can be proven safe in real-valued \dL{}. %

In practice, real-valued arithmetic does not exist, and finite precision introduces pitfalls: When $p$ is close to $0$, even infinitesimal perturbations to $v$ can cause $p < 0$, leading to a collision. Braking earlier \emph{cannot} resolve this, as similar perturbations could, by induction, still push the robot into an unsafe region.
This illustrates a \emph{non-robust} control envelope, highlighting the need to \emph{rethink} the system structure---or at least the model---by, e.g., also admitting negative velocities.

\paragraph{Our Approach}
\looseness=-1
Our work extends \dL{}-based safety guarantees to finite-precision NNs while:
\begin{enumerate*}
    \item still providing guarantees w.r.t. precise real-valued physical dynamics modeled in \dL{} and
    \item leveraging efficient, state-of-the-art real-valued NN verification tools.
\end{enumerate*}
We build on VerSAILLE~\cite{Teuber2024}, which proves real arithmetic refinement relations between NNs and provably safe \dL{} control envelopes.
Just as VerSAILLE, and as indicated in \Cref{fig:overview}, our approach assumes the availability of a safe $\dL{}$ model of the analyzed control system.
In particular, this \dL{} model comprises a (nondeterministic) \emph{control envelope} which encodes provably safe control actions.
To consider perturbations arising from sensing, computation (e.g., fixed-point arithmetic), or actuation, we model such perturbations as real-valued hybrid programs, termed as \emph{angelic perturbations}.
For example, in our robot example, we define an angelic perturbation $\robotSimplePerturb$ executed after the original control envelope chooses $v$:
It nondeterministically selects a bounded perturbation $\left|\varepsilon_v\right| \leq \delta_v$ and then updates $v$ via $v \coloneqq v + \varepsilon_v$.
This captures bounded perturbation of up to $\delta_v$ for the robot's actuation, whether due to numerical imprecision or actuator inaccuracies.

As can be seen in \Cref{fig:overview}, these angelic perturbations can be used to multiple ends:
First, we can use them to analyze the robustness of a given \dL{} model/control envelope.
To this end, we provide a game-based robustness definition and decidable sufficient criteria for robustness.
Second, for a robust control envelope, we extend VerSAILLE's analysis to take account of angelic perturbations in a decidable manner.
This allows us to prove infinite-time horizon NNCS safety under perturbation.
Finally, for NNs that are safe under perturbation, we can leverage sound quantization techniques, which synthesize a fixed-point implementation of the real-valued NN where the maximal error bound matches our angelic perturbation.

\emph{Towards our contribution (C1)}, 
we provide a game-theoretic definition for the robustness of control envelopes:
There must \emph{exist} control actions that keep the system safe irrespective of the environment's \emph{angelic perturbations}.
Classic \dL{} safety properties ($\left[\alpha\right]\psi$) interpreted as game
show that Demon, who has no influence on execution, has a winning strategy irrespective of Angel's control and environment choices.
Robustness hands over the controller choices to a \emph{good} Demon who must choose actions to weather a \emph{bad} Angel
adversarially choosing perturbations and environment behavior.
Our formulation induces \emph{decidable} real-arithmetic conditions applicable to \emph{any} discrete, loop-free angelic perturbation.
We identify \emph{liveness} (a control envelope does not get \emph{stuck}). as a special case of robustness. %
For our robot, a control envelope with $v \geq 0$ is not robust under $\robotSimplePerturb$, but an envelope with bidirectional movement is (choose any $v$ such that $p - T(v+\delta) \geq 0$).

\looseness=-1
\emph{Concerning (C2)}, we extend VerSAILLE %
to account for angelic perturbations: %
NN verification reduces to checking whether the NN implements a \emph{winning strategy} for the robustness game formulated in (C1)
and robustness becomes a \emph{necessary} condition for \dL{}-based NN verification under finite precision.
For the robot with bidirectional movement, the real-valued condition $p \geq 0 \rightarrow p + T f(p) \geq 0$, where $f\left(p\right)$ is the velocity computed by the NN $f$ can be extended to $-\delta_{v} \leq \varepsilon_{v} \leq \delta_{v} \land p \geq 0 \rightarrow p + T(f\left(p\right)+\varepsilon_{v}) \geq 0$, which is also compatible with existing linear and nonlinear NN verification tools.
Disproving NN safety amounts to finding a concrete input $p$ and perturbation $\varepsilon_v$ violating the specification. Proving the specification yields an \emph{infinite-time horizon safety guarantee} for the NN control system under bounded perturbations.
Our approach systematically handles a broad class of perturbations with decidable construction from \dL{} control envelopes.

\emph{Addressing (C3)}, we leverage NN safety guarantees under bounded perturbations
and utilize Daisy for mixed-precision fixed-point tuning, with added pre- and post-processing steps to support typical NN structures and operations, enabling direct hardware synthesis using Xilinx Vivado~\cite{websiteOfVivado}. 

Thus, our \emph{end-to-end} approach ensures infinite-horizon safety under bounded finite-precision errors and produces sound, efficient, hardware-deployable NN controllers.
While in this paper, we focus on NNs due to current tool support, the approach is broadly applicable and can be straightforwardly extended to other types of controllers.

\section{Background}\label{background}
 This section provides background on proving infinite-time horizon safety for NN Control Systems using \dL{} (\Cref{subsec:background:safety_verification}) and on efficient code generation for finite-precision (\Cref{subsec:background:efficient_code}).
 We illustrate these concepts using the robot example from ~\Cref{fig:running:robot_idea}.

\subsection{Safety Verification}
\label{subsec:background:safety_verification}
We review Differential Dynamic Logic~\cite{DBLP:journals/jar/Platzer08} for abstractly reasoning about the hybrid (continuous and discrete) behavior of Cyber-Physical Systems, Differential Game Logic~\cite{dGL} for handling adversarial dynamics, and describe the \dL{}-based NN verification technology VerSAILLE~\cite{Teuber2024}.
\subsubsection{Differential Dynamic Logic}
\label{subsec:dl}
Differential Dynamic Logic (\dL)~\cite{DBLP:journals/jar/Platzer17,Platzer2020,Platzer2012,DBLP:journals/jar/Platzer08} is a (first-order) dynamic logic~\cite{DBLP:books/sp/Harel79,harel-dynamic-logic}, i.e., a modal logic over a state space $\stateSpace$ whose modalities are parametrized with programs.
The syntax of \emph{hybrid programs} analyzed in \dL{} is given by the following grammar, where $x$ is a variable, $Q$ is a real arithmetic formula, $e$ is either $*$ or a real arithmetic term,
and $f\left(x\right)$ is a real arithmetic function over $x$ (with $\alpha$ and $\beta$ representing hybrid programs):
$
\scalebox{0.915}{$
\alpha, \beta \Coloneqq~
\underbrace{x'=f(x) \; \& \; Q}_{\text{continuous}} %
\;\mid\; \underbrace{%
x \coloneqq e 
\;\mid\; ?(Q) 
\;\mid\; \alpha;\beta 
\;\mid \alpha \cup \beta%
}_{\text{discrete and loop-free}}\;\mid\; 
\underbrace{(\alpha)^* }_{\text{loop}}
$}
$

The semantics of hybrid programs are described as state transitions where a state $\stateOne$ is a map from variables $x$ to (real) values $\stateOne\left(x\right)$.
For example, for $e$ a real arithmetic term, the semantics of variable assignment are defined as 
$
\llbracket x \coloneqq e \rrbracket = 
\left\{
\left(\stateOne,\stateTwo\right) \in \stateSpace^2 ~\middle|~
\stateTwo = \stateOne_x^e
\right\}
$
where $\stateOne_x^e$ is a state that evaluates $x$ to $e$ and all other variables $y$ to $\stateOne\left(y\right)$.

The defining feature of hybrid programs is their ability to model \emph{continuous} dynamics:
$x'=f\left(x\right)\&Q$ denotes a program where
$x$ nondeterministically evolves along the given differential equation within an evolution constraint $Q$, allowing $x$ to take \emph{any} reachable value satisfying $Q$.
The primitive ${x\coloneqq *}$ represents a nondeterministic assignment.
The remaining constructs are: predicate checks (no transition if the check fails), sequential composition, nondeterministic choice, and looping. 
Safety properties can be expressed as formulas $P \implies \left[\alpha\right] Q$, stating that, if the precondition $P$ holds, then $Q$ holds after all (nondeterministic) executions of the hybrid program $\alpha$.
For example, the formula $x=2 \implies \left[y \coloneqq x \cup y \coloneqq x+x\right] y>0$ is valid because for any state with $x=2$, all possible executions ensure that $y>0$ (specifically $y\in\left\{2,4\right\}$) after execution.

\dL{} also provides a dual modality to express liveness: $P \implies\langle \alpha \rangle Q$ guarantees that, starting from any state satisfying 
$P$, it is \emph{possible} to reach \emph{some} state where 
$Q$ holds.
Moreover, the equivalence $\left[\alpha\right]Q \equiv \neg \langle \alpha\rangle \neg Q$ applies.

\looseness=-1
We also use the subset of \emph{concrete, discrete, and loop-free} programs, corresponding to the discrete and loop-free production rules from the hybrid program grammar above.
Vectors of variables are denoted as $\bar{x}$ while $x$ is an individual variable.

Safety guarantees in \dL{} typically have the form: 
\begin{equation}
\label[formula]{fml:safety_predicate}
\safetyPredicate{\precondition,\controllerProgram,\environmentProgram,\postcondition} \equiv
\precondition\implies\left[\left(\controllerProgram;\environmentProgram\right)^*\right] \postcondition,
\end{equation}
where $\precondition$ defines initial conditions and $\postcondition$ the safety guarantee; $\controllerProgram$ and $\environmentProgram$ describe the (nondeterministic) \emph{control envelope} and the environment, respectively.

\subsubsection{Differential Game Logic}
\looseness=-1
Differential Game Logic (\dGL{}) extends \dL{} to reason about \emph{hybrid games}, i.e., hybrid, noncooperative zero-sum games between two players, \emph{Angel} and \emph{Demon}.
A game is expressed as a program $\alpha$ in a syntax extending the hybrid program grammar
as summarized below, however, for a comprehensive introduction, see~\cite{dGL}. 

For a game $\alpha$, $\left[\alpha\right] Q$ states that Demon has a winning strategy to achieve $Q$, while $\langle\alpha\rangle Q$ indicates that Angel has a winning strategy.
In the \dL{} primitives, we assume Angel controls \emph{nondeterministic decisions} by default, with control explicitly transferred to Demon via additional primitive $\alpha^d$.

\looseness=-1
In this interpretation, a \dL{} safety property $P \implies \left[\alpha\right]Q$ for a classic hybrid program (i.e., without duality) states that, when starting from $P$, Demon has a winning strategy to ensure $Q$ regardless of Angel's nondeterministic decisions.
For example, the game $\alpha\equiv\left(v \coloneqq 1 \cup v \coloneqq -1\right)^d;x'=v$ states that Demon can choose $v\in\left\{-1,1\right\}$ (observe the duality operator) and subsequently Angel controls for how long to evolve $x$ along the differential equation.
The formula $x \neq 0 \implies \left[\alpha\right] x \neq 0$ is then valid:
Demon can choose $v$ such that, even as Angel controls the evolution of $x$, $x$ never reaches $0$. 
Specifically, Demon selects $v \coloneqq 1$ iff $x > 0$, ensuring $x\neq 0$. Hence, Demon has a winning strategy for $x \neq 0$.
Although \dGL{} has a substantially different semantical foundation~\cite{dGL}, it alligns with \dL{} semantics for programs and formulas without duality~\cite[p. 520ff]{Platzer2018}.

\subsubsection{Verifying NN with VerSAILLE}
\label{subsec:modelplex}
As indicated in \Cref{subsec:dl}, \dL{} is used to prove the safety of abstract control envelopes, typically formulated as in \Cref{fml:safety_predicate}.
For real-world systems, however, we must relate these abstract safety proofs to concrete controller implementations outside \dL{}.
This requires showing that the idealized implementation \emph{refines} the safe control envelope ~\cite{DBLP:conf/lics/LoosP16,DBLP:conf/ijcar/PrebetP24}, i.e., that \emph{all} behaviors of the idealized implementation are also possible in the \dL{} model~\cite{DBLP:conf/lics/LoosP16,DBLP:conf/ijcar/PrebetP24}.
To this end, ModelPlex~\cite{DBLP:journals/fmsd/MitschP16}
can derive correct-by-construction \emph{controller monitor formulas}.
Given a control envelope $\controllerProgram$, its monitor formula $\controllerMonitor$ is a first-order formula over pre- and post-states, satisfied only if a state transition is modeled by $\controllerProgram$.
Formally:
\begin{definition}[Correct Controller Monitor~\cite{DBLP:journals/fmsd/MitschP16}]
\label{def:correct_monitor}
A \emph{controller monitor formula} $\controllerMonitor$ is \emph{correct} for a hybrid program controller $\controllerProgram$ with bound variables $x_1,\dots,x_n$ iff the following \dL{} formula is valid:
$
\controllerMonitor \implies \langle \controllerProgram \rangle \bigwedge_{i=1}^n x_i = x_i^+
$.
\end{definition}
At least, $\controllerMonitor$ typically includes the variables $x_1,x_1^+,\dots,x_n,x_n^+$ describing the bound variables of \controllerProgram{} before and after the execution of \controllerProgram{}. %

Originally used for constructing monitors and shields~\cite{DBLP:conf/aaai/FultonP18,DBLP:conf/pldi/BohrerTMMP18},
 ModelPlex was recently shown to support a priori safety verification through VerSAILLE~\cite{Teuber2024}.
Given a safe \dL{} model (\Cref{fml:safety_predicate}) and a loop invariant $\stateFormula$, ModelPlex derives the controller monitor formula $\controllerMonitor$.
We model the NN's behavior as a before-after predicate $\implementationPredicate{\bar{x}}{\bar{x}^+}$.
 Verifying infinite-horizon safety of the NNCS then reduces to proving the following real arithmetic formula~\cite[Thm. 1]{Teuber2024}:
\begin{equation}
\label[formula]{fml:versaille_fol_obligation}
\stateFormula \land \implementationPredicate{\bar{x}}{\bar{x}^+} \rightarrow \stateFormula
\end{equation}
The obtained guarantee has a rigorous foundation in \dL{}: 
Based on $\implementationPredicate{x}{x^+}$, we derive a hybrid program (called \emph{nondeterministic mirror}, denoted $\alpha_{\text{refl}}\left(\implementationPredicateSym{}\right)$) mirroring the behavior of the before-after predicate \emph{within} \dL{}.
Proving \Cref{fml:versaille_fol_obligation} then implies safety when replacing \controllerProgram{} with $\alpha_{\text{refl}}\left(\implementationPredicateSym{}\right)$.

\subsection{Efficient Code Generation}
\label{subsec:background:efficient_code}
We review the background on fixed-point arithmetic and mixed-precision tuning here.

\subsubsection{Fixed-Point Arithmetic}
Fixed-point numbers represent values as integers with an (implicit) format $\langle s, Q, \pi \rangle$, where $s \in {0, 1}$ indicates the presence of a sign bit, $Q \in \mathbb{N}$ is the total word length, and $\pi \in \mathbb{N}$ specifies the binary point position (counted from the least significant bit). Bit allocation defines an integer part, $I$, determining the range [$-2^I, 2^I$], and a fractional part, $\pi$, controlling the \emph{precision}: more fractional bits yield higher precision. Assuming truncation as the rounding mode (default for synthesis tools like Xilinx Vivado~\cite{websiteOfVivado}), the maximum roundoff error is $2^{-\pi}$. Fixed-point operations use efficient integer arithmetic and bit-shifting~\cite{anta2010}, making them ideal for resource-constrained systems.

\subsubsection{Mixed-Precision Tuning}\label{background:tuning}
Mixed-precision tuning assigns different bitwidths to variables and constants to optimize resource usage while satisfying error bounds. Several techniques~\cite{Chiang2017,damouche2018mixed,Daisy,bessai2022fixed} have been proposed to automate this process. In this work, we focus on the tool Daisy~\cite{Daisy} and briefly summarize its approach below.

Daisy~\cite{Daisy} performs mixed-precision tuning by heuristically navigating the search space using \emph{delta debugging}~\cite {DBLP:conf/esec/Zeller99}. To soundly compute roundoff errors during the search, Daisy applies either interval or affine arithmetic, depending on the configuration. The process starts from a minimal bit length, uniformly increasing the precision to meet the error target, and then recursively partitioning variables to selectively lower bitwidths. Among multiple feasible assignments, Daisy selects the one minimizing an \emph{area-based} cost function. The resulting mixed-precision fixed-point code can be compiled by synthesis tools like the Xilinx HLS compiler. Daisy is most effective for straight-line numerical code without complex control structures such as loops, conditionals, or arrays.

\section{Safety under Perturbation}
As discussed in \Cref{subsec:background:safety_verification}, VerSAILLE's safety guarantees rely on idealized NN implementations in real arithmetic. This has the advantage of enabling
powerful real-arithmetic solvers~\cite{DBLP:conf/tacas/MouraB08,DBLP:conf/cav/BjornerN24,DBLP:conf/cade/KremerRBT22,Mathematica} and real-valued NN verification tools (as reported in recent surveys and competitions~\cite{DBLP:journals/jmlr/0005BHR24,DBLP:journals/sttt/BrixMBJL23,DBLP:journals/corr/abs-2412-19985}).
However, it raises the question of whether these safety results hold when the NN is implemented in fixed-point arithmetic. In practice,
real-world systems deviate from their idealized real-valued models along three dimensions:
\begin{enumerate}[label=(I\arabic*)]
    \item \emph{Sensor Readouts:} 
    Limited bandwidth or finite measurement accuracy affects input precision.
    \item \emph{Roundoff Errors:} %
    Errors can arise during computations.
    \item \emph{Actuator Limitations:} Physical constraints may impact the precision of chosen control actions.
\end{enumerate}

In principle, such factors can be mitigated by designing control envelopes for environments that explicitly model sensor and actuator noises, and computation errors.
For instance, the hybrid program $\robotSimplePerturb \equiv (\varepsilon_v \coloneqq *;?\left(\left|\varepsilon_v\right|\leq \delta_v\right);v \coloneqq v + \varepsilon_v)$
models bounded perturbations in computing $v$ in the robot example from \Cref{subsec:methods:example}.
However, many case studies, e.g., ~\cite{DBLP:journals/sttt/JeanninGKSGMP17, DBLP:conf/fmics/WuRF20, DBLP:journals/tiv/SelvarajAF23,Teuber2024}, only assume ideal sensing, actuation, and computation for simplicity.
This is unsurprising as hybrid system analysis is already complex 
(even undecidable~\cite{DBLP:conf/hybrid/PlatzerC07}) even without accounting for finite-precision effects.

This work proposes a technique to check %
control envelope robustness to sensing, actuation, and roundoff errors, given a safe control envelope oblivious to such perturbations.
We further show how these control envelopes can be used to verify fixed-point implementations.

We begin by recalling how VerSAILLE guarantees safety for an idealized implementation for the robot example (\Cref{subsec:methods:example}). Building on this, our contribution starts with the formalization of perturbations (\Cref{subsec:method:perturbation}), a robustness property for control envelopes and decidable conditions for proving it (\Cref{subsec:method:robustness}). We then extend VerSAILLE to support verification under perturbation (\Cref{subsec:method:verification}). While our method is illustrated using the running example --- \ifextension{with some of the formal notions deferred to \Cref{apx:running-example}}{with some formal notions deferred to the extended version} --- 
\Cref{sec:evaluation} shows its applicability across realistic case studies.

\subsection{Setup: Idealized Implementation Safety with VerSAILLE}
\label{subsec:methods:example}
\begin{figure}[t]
    \centering
    \begin{align*}
    \envelopeExactOneDirOneWall~&\equiv~\rlap{$
    \left(v \coloneqq *; ?\left(0 \leq v \leq V_{\text{max}} \land Tv \leq p\right)\right)$}
    \\
    \envelopeExactTwoDirOneWall~&
    \equiv~\rlap{$\left(v \coloneqq *; ?\left(-V_{\text{max}} \leq v \leq V_{\text{max}} \land 0 \leq p-Tv\right)\right)$}
    \\
    \envelopeExactTwoDirTwoWall~&
    \equiv~\rlap{$\left(v \coloneqq *; ?\left(-V_{\text{max}} \leq v \leq V_{\text{max}} \land 0 \leq p-Tv \leq W \right)\right)$}  \\
    \preExactOneWall~&\equiv~ p, V_{\text{max}} \geq 0~\land~T>0
    &
    \postExactOneWall~
    &\equiv~ 0 \leq p\\
    \preExactTwoWall~&\equiv~ \preExactOneWall~\land~ p \leq W > 0 &
    \postExactTwoWall~
    &\equiv~ 0 \leq p \leq W
    \end{align*}
    \caption{Possible control envelopes, pre- and postconditions for robot (\Cref{overview})}
    \label{fig:controller_safety}
\vspace*{-0.5cm}
\end{figure}

We recall how \dL{} and VerSAILLE can derive safety guarantees for an idealized implementation of our robot example from \Cref{fig:running:robot_idea}.
The system state is given by position $p$ and velocity $v$ along with a clock variable $t$ tracking elapsed time.
We formalize the robot's physical behavior ($p'=v$), or \emph{plant model}, as a hybrid program $\environmentProgram$ ensuring control actions occur at least every $T$ seconds
(\ifextension{for more details, see \Cref{apx:running-example}}{see details in extended version}).

Different control scenarios and corresponding pre/postconditions and control envelopes are shown in \Cref{fig:controller_safety}.
In the single-wall case, the safety property $\postExactOneWall$ is guaranteed by the envelopes $\envelopeExactOneDirOneWall$ and $\envelopeExactTwoDirOneWall$, which differ in whether bidirectional motion is allowed. For the two-wall case, the robot must stay within bounds defined by $\postExactTwoWall$, requiring a different control envelope $\envelopeExactTwoDirTwoWall$.

With suitable loop invariants, we can prove safety for $\envelopeExactOneDirOneWall$ and $\envelopeExactTwoDirOneWall$ w.r.t. $\preExactOneWall$ and $\postExactOneWall$, as well as safety for $\envelopeExactTwoDirTwoWall$ w.r.t. $\preExactTwoWall$ and $\postExactTwoWall$ in the theorem prover KeYmaera~X~\cite{Fulton2015} 
(\ifextension{proof in \Cref{apx:running-example}}{proof in extended version}).
The envelope $\envelopeExactOneDirOneWall{}$ slightly deviates from the controller in~\cite{DBLP:conf/pldi/BohrerTMMP18}, which checks for $TV \leq p$ (instead of $Tv \leq p$).
This change makes the envelope more \emph{permissive} by allowing intermediate velocities $0 < v < V_{\text{max}}$ when choosing $v=V_{\text{max}}$ would no longer be safe; however, $\envelopeExactOneDirOneWall{}$'s behavior directly depends on the relation between the (changing) variables $p$, $v$ and the computated $Tv$.

We can now use the \dL{} safety results to verify an idealized NN implementations. For demonstration, we assume the idealized implementation is described by the predicate $\implementationPredicateRobotExact{p}{v^+}\equiv \big(v^+ = \frac{-1}{0.01\left(p+10\right)}+M\big)$.
Infinite-time horizon safety for one-wall, unidirectional case then be verified by instantiating \Cref{fml:versaille_fol_obligation} and
 proving the following:
\vspace*{-1mm}
$$
\smash{\underbrace{p \geq 0}_{\stateExactOneDirOneWall}} \land
\implementationPredicateRobotExact{p}{v^+}
\implies
\smash{\underbrace{(0 \leq v^+ \leq V_{\text{max}} \land 0 \leq p - Tv^+)}_{\monitorExactOneDirOneWall}}
\vspace{5mm}
$$
where $\stateExactOneDirOneWall$ is the loop invariant and $\monitorExactOneDirOneWall$ is the controller monitor derived from $\envelopeExactOneDirOneWall$ via ModelPlex.
In practice, this validity is not proven directly by  \Cref{fml:versaille_fol_obligation}, but via highly tuned NN verification tools that implicitly reason about \implementationPredicateSym.

For $M=V_{\text{max}}=10$, we prove validity of the formula using an SMT solver.
Let $\alpha_{\text{refl}}\left(\implementationPredicateRobotExactSym\right)$ be the nondeterministic mirror of $\implementationPredicateRobotExactSym$, i.e., the hybrid program capturing its behavior.
This implies the safety predicate $\safetyPredicate{\preExactOneWall,\alpha_{\text{refl}}\left(\implementationPredicateRobotExactSym\right),\environmentProgram,\postExactOneWall}$.
Using the same approach for $\envelopeExactTwoDirTwoWall$,
we also prove safety for $\implementationPredicateRobotExact{p}{v^+}$ for $M=V_{\text{max}}=10,W=100$.
However, VerSAILLE's analysis assumes real arithmetic.
Hence, it ignores roundoff errors introduced by realistic, fixed-point implementations of $\implementationPredicateRobotExactSym$.

\subsection{Formalizing Perturbations}
\label{subsec:method:perturbation}
We formalize perturbations as concrete, discrete, loop-free hybrid programs.
As seen earlier, $\robotSimplePerturb$ captures \emph{post}-controller perturbations to $v$, which may represent roundoff errors, actuation errors, or both (with $\delta_v$ as maximal roundoff, maximal deviation in actuation, or the combined bound). However, sensor perturbations require modifying the state \emph{before} controller execution.  
To cover all perturbation profiles (I1-I3), we formalize general perturbation as a tuple $\left(\alpha_{\text{pre}},\alpha_{\text{post}}\right)$,
 where $\alpha_{\text{pre}}$ applies before the controller and $\alpha_{\text{post}}$ after. We call such pairs \emph{angelic perturbations}:%
\begin{definition}[Angelic Perturbation]
\label{def:angelicPerturbation}
An \emph{angelic perturbation} consists of two discrete, concrete, loop-free hybrid programs $\alpha_{\text{pre}},\alpha_{\text{post}}$. We denote an angelic perturbation as a tuple (i.e. $\left(\alpha_{\text{pre}},\alpha_{\text{post}}\right)$).
\end{definition}%
\noindent
In the robot example, we consider two such perturbation profiles:
\begin{align*}
    \preProgramParam{1}\enspace\equiv&
    \enspace?\left(\top\right) &
    \postProgramParam{1}\enspace&
    \equiv\enspace\robotSimplePerturb\hspace*{2cm}\\
    \preProgramParam{2}\enspace\equiv&
    \rlap{$\enspace p^- \coloneqq p; \varepsilon_p\coloneqq*;?\left(\left|\varepsilon_p\right|\leq\delta_p\right);p\coloneqq p+\varepsilon_p$}\\
    \postProgramParam{2}\enspace\equiv&
    \rlap{$\enspace
    p\coloneqq p^-;\varepsilon_v\coloneqq *; ?\left(\left|\varepsilon_v\right| \leq \delta_v\right); v \coloneqq v + \varepsilon_v$}
\end{align*}
Here, $(\preProgramParam{1}, \postProgramParam{1})$ perturbs only $v$ by at max $\delta_v$; whereas, $(\preProgramParam{2}, \postProgramParam{2})$ perturbs both the sensor input $p$ without affecting the system's dynamics (the original value is restored after control execution) and the control output $v$.

\subsection{Envelope Robustness}
\label{subsec:method:robustness}
Before verifying implementations, we must check if the envelope itself is robust to angelic perturbations
---a necessary condition for subsequent verification as we use the envelope to check if our implementations represent a winning strategy.

We formulate the robustness of control envelopes as a game: rather than requiring that \emph{all} control actions remain safe under perturbation, we require that for any state, \emph{there exists} a control action that
\begin{enumerate*}
    \item is permitted according to the envelope, and
    \item guarantees safety even when perturbed by the given angelic perturbation.
\end{enumerate*}
The \emph{good} Demon, choosing an action from within the envelope, hence must take into account the perturbations of the \emph{bad} Angel.

To check this property, we formulate a \dGL{} game
where Demon chooses from \controllerProgram{} and Angel controls perturbations and the environment.
Robustness (in \dGL{}) is then defined as Demon having a winning strategy to guarantee $\postcondition$:
\begin{definition}[Robustness]
\label{def:robust_control_envelopes}
\looseness=-1
A control envelope $\controllerProgram$ with a safety guarantee 
$\safetyPredicate{\precondition,\controllerProgram,\environmentProgram,\postcondition}$ is \emph{robust} to angelic perturbations $\left(\preProgram,\postProgram\right)$ iff the following is valid:
\begin{equation}
\label{fml:dgl_safety_abstract}
\precondition \implies \left[\left(
\preProgram;\left(\controllerProgram\right)^d;\postProgram;\environmentProgram
\right)^*\right]\postcondition
\end{equation}    
\end{definition}

\Cref{def:robust_control_envelopes} assumes that
\controllerProgram{} already has a real-valued safety guarantee (e.g.\ from~\cite{DBLP:journals/sttt/JeanninGKSGMP17,DBLP:conf/fmics/WuRF20,DBLP:journals/tiv/SelvarajAF23,Teuber2024}).
We \emph{reuse} this knowledge (and the safety proof) to check the robustness of \controllerProgram{} utilizing insights on the decidability of differential refinements~\cite{DBLP:conf/ijcar/PrebetP24}.
We propose a practical, \emph{sufficient} criterion for controller robustness that can be checked in decidable real arithmetic
(\ifextension{see proof on \cpageref{proof:robust_envelope}, further simplification see \Cref{apx:simplified_conditions}}{see extended version for proofs and additional simplifications}).

\begin{textAtEnd}
    To prove \Cref{thm:robust_envelope} we require the \emph{Pullback Axiom} which we prove to be correct here:
    \end{textAtEnd}
    \begin{lemmaE}[Pullback Axiom][all end]
        \label{lem:pullback}
        The \emph{Pullback Axiom} is sound and derivable in \dL{}'s proof calculus:
        \begin{itemize}
            \item[\ensuremath{P}]
            $
            \forall \bar{x}^+
            \left(\psi \implies \langle \alpha\rangle \xi\right)
            \land
            \exists \bar{x}^+
            \left(
            \psi \land
            \forall \bar{x}
            \left(\xi \implies \phi\right)
            \right)
            \implies
            \left(
            \langle \alpha \rangle \phi
            \right)
            $
        \hfill
        $
        \begin{array}{ll}
             \bar{x} &\supseteq V\left(\langle\alpha\rangle \phi\right)\\
        \bar{x}^+ &= FV\left(\psi\land\xi\right) \setminus \bar{x}
        \end{array}
        $
        \end{itemize}
        Where $FV\left(\cdot\right)$ is the set of free variables of a formula/program and $V\left(\cdot\right)$ is the set of all variables appearing in a formula/program.
    \end{lemmaE}
    \begin{proofE}
    The proof is based on the $K_{\langle\rangle}$ axiom derived in \cite[Appendix B.4]{Platzer2012}:
    \begin{itemize}
            \item[\ensuremath{K_{\langle\rangle}}] $\left[\alpha\right] \left(\psi \implies \phi\right) \implies \left(\langle \alpha\rangle \psi \implies \langle \alpha\rangle \phi\right)$
    \end{itemize}
    Using this axiom, we can derive the axiom above:
    Starting from 
    $\forall \bar{x}^+
    \left(\psi \implies \langle \alpha \rangle \xi\right)$
    and
    $
    \exists \bar{x}^+
    \left(
    \psi \land
    \forall \bar{x}
    \left(\xi \implies \phi\right)
    \right)$
    we can derive via first-order reasoning that this implies
    $
    \exists \bar{x}^+
    \left(
    \psi \land
    \forall \bar{x}
    \left(\xi \implies \phi\right)
    \land
    \left(\psi \implies \langle \alpha \rangle \xi\right)
    \right)
    $
    via local Modus Ponens, this can be turned into
    $
    \exists \bar{x}^+
    \left(
    \forall \bar{x}
    \left(\xi \implies \phi\right)
    \land
    \langle \alpha \rangle \xi
    \right)
    $.
    Via $V$ (all bounded variables of $\alpha$ are not free in the formula), we get that this implies
    $
    \exists \bar{x}^+
    \left(
    \left[\alpha\right]
    \forall \bar{x}
    \left(\xi \implies \phi\right)
    \land
    \langle \alpha \rangle \xi
    \right)
    $.
    Instantiating the variables of $\bar{x}$ with their outside version this implies
    $
    \exists \bar{x}^+
    \left(
    \left[\alpha\right]
    \left(\xi \implies \phi\right)
    \land
    \langle \alpha \rangle \xi
    \right)
    $.
    Applying $K_{\langle\rangle}$ this implies
    $
    \exists \bar{x}^+
    \left(
    \left(\langle\alpha\rangle\xi \implies \langle\alpha\rangle\phi\right)
    \land
    \langle \alpha \rangle \xi
    \right)
    $.
    Via local Modus Ponens, this implies
    $\exists \bar{x}^+\langle\alpha\rangle\phi$.
    Since $\bar{x}^+$ is not part of $\phi$ or $\alpha$, this implies
    $\langle\alpha\rangle\phi$.
    This concludes the proof.
\end{proofE}
\begin{theoremE}[Robustness of Control Envelopes][end,restate,text link=]
\label{thm:robust_envelope}
Let \controllerProgram{} be as in \Cref{def:robust_control_envelopes} and $\stateFormula$ be the inductive invariant used to prove $\safetyPredicateSym\left(\dots\right)$.
For angelic perturbation $\preProgram,\postProgram$ and monitor formulas $\controllerMonitor{},\preMonitor{},\postMonitor{}$ for $\controllerProgram{},\preProgram{},\postProgram{}$, the validity of the following real arithmetic formula implies the validity of \Cref{fml:dgl_safety_abstract}.
\begin{align}
\forall \bar{x}_0&~\forall{x}_1~\exists \bar{x}_2~\forall \bar{x}_3\enspace
\big(\stateFormula\left(\bar{x}_0\right) \land \preMonitor\left(\bar{x}_0,\bar{x}_1\right)
~\implies\nonumber\\
&\left(
\controllerMonitor\left(\bar{x}_1,\bar{x}_2\right) \land
\left(
\postMonitor\left(\bar{x}_2,\bar{x}_3\right)
\implies
\controllerMonitor\left(\bar{x}_0,\bar{x}_3\right)
\right)
\right)\big)\label[formula]{fml:arithmetic_winning_strategy}
\end{align}
\end{theoremE}
\begin{proofE}
\label{proof:robust_envelope}
For our proof, we assume the validity of \Cref{fml:arithmetic_winning_strategy}.
We now begin a proof of \Cref{fml:dgl_safety_abstract} via the \dGL{} proof calculus~\cite{dGL}.
First, we apply the loop invariant rule for \stateFormula{} to \Cref{fml:dgl_safety_abstract}.
The formulas $\precondition \implies \stateFormula$ and $\stateFormula \implies \postcondition$ immediately follow from our assumption that \Cref{fml:safety_predicate} was proven via the invariant \stateFormula{}.
From this assumption, we also know that:
\[
\stateFormula \implies \left[\controllerProgram;\environmentProgram\right] \stateFormula.
\]
It remains to show that:
\[
\stateFormula \implies \left[\preProgram;\left(\controllerProgram\right)^d;\postProgram;\environmentProgram\right] \stateFormula.
\]
Let $V$ be the set of all variables in the above formula.
For each variable $x\in V$ part of this formula, we now add a fresh, discrete ghost variable $x^-$, assigned to the same value as $x$.
Simple reasoning then yields that the above formula can be shown by showing the validity of the formula\footnotemark{}
\footnotetext{
Where $\bar{x}^- = \bar{x}\enspace\equiv\enspace\bigwedge_{x \in V} x^- = x$
}
\[
\stateFormula \land \stateFormula^- \land \bar{x}^- = \bar{x} \implies \left[\preProgram;\left(\controllerProgram\right)^d;\postProgram;\environmentProgram\right] \stateFormula
\]
where $\stateFormula^-$ is the formula $\stateFormula$ where each apperance of $x \in V$ has been replaced by $x^-$.
Via the monotonicity rule, we can decompose this into two proof obligations from hereon referred to as (A) and (B):
\paragraph*{(A)}
The first proof obligation reads as follows:
\begin{align*}
\stateFormula& \land \stateFormula^- \land \bar{x}^- = \bar{x} \implies \\
&\left[\preProgram;\left(\controllerProgram\right)^d;\postProgram\right]
\left(\stateFormula^- \land \controllerMonitor\left(\bar{x}^-,\bar{x}\right)\right).
\end{align*}
Via application of the $\left[;\right]$ and $\left[^d\right]$ rules, this can be transformed into the following formula:
\begin{align*}
\stateFormula& \land \stateFormula^- \land \bar{x}^- = \bar{x} \implies \\
&\left[\preProgram\right]\langle \controllerProgram\rangle \left[\postProgram\right]
\left(\stateFormula^- \land \controllerMonitor\left(\bar{x}^-,\bar{x}\right)\right).
\end{align*}
Importantly, this formula no longer contains the dual operator.
Hence, for this formula, the state transition semantics of \dL{} and the winning region semantics of \dGL{} coincide~\cite[p. 520ff]{Platzer2018}.
As ModelPlex's guarantees are derived w.r.t \dL{} semantics, we, therefore, prove this statement w.r.t. \dL{} semantics knowing that \dGL{} validity follows from \dL{} validity for duality-free formulas.
We can focus on proving the validity of the following \dL{} formula:
\begin{align*}
\stateFormula^-& \land \bar{x}^- = \bar{x} \implies \\
&\left[\preProgram\right]\langle \controllerProgram\rangle \left[\postProgram\right]
\left(\controllerMonitor\left(\bar{x}^-,\bar{x}\right)\land\stateFormula^-\right).
\end{align*}
Observe that we assume $\preProgram,\postProgram$ are concrete, discrete, loop-free hybrid programs. Hence, as observed by Prebet and Platzer~\cite{DBLP:conf/ijcar/PrebetP24}, the monitoring formulas generated by ModelPlex are exact, that is, e.g. for $\preProgram$ we have $\preMonitor\left(\bar{x},\bar{x}^+\right) \iff \langle \preProgram\rangle \bar{x}=\bar{x}^+$.
Hence we have
$\stateFormula^- \land \bar{x}^- = \bar{x}\implies\left[\preProgram\right]\left(\stateFormula^- \land \preMonitor\left(\bar{x}^-,\bar{x}\right)\right)$.
Via a Monotonicity argument, we can then remove $\preProgram$ from the previous proof obligation. Using fresh, discrete ghost variables we copy the values of $\bar{x}$ into a new variable vector $\bar{x}_1$ which yields
\begin{align*}
\stateFormula^-& \land \preMonitor\left(\bar{x}^-,\bar{x}_1\right)
\land \bar{x}_1 = \bar{x}
\implies\\
&\langle \controllerProgram\rangle \left[\postProgram\right]
\left(\controllerMonitor\left(\bar{x}^-,\bar{x}\right)\land\stateFormula^-\right).
\end{align*}
We can now make use of the Pullback Axiom (see \Cref{lem:pullback}), which allows us to pull the state after executing $\controllerProgram$ into the current state.
To this end, we instantiate $\forall \bar{x}^+ \left(\psi \implies \langle\alpha\rangle\xi\right)$ with $\forall \bar{x}^+ \left(\controllerMonitor\left(\bar{x},\bar{x}^+\right) \implies \langle \controllerProgram\rangle \bar{x}=\bar{x}^+\right)$ which is valid by construction.
Instantiating $\phi$ with $\left[\postProgram\right]
\left(\controllerMonitor\left(\bar{x}^-,\bar{x}\right)\land\stateFormula^-\right)$ and simplifying $\bar{x}$ to $\bar{x}_1$, we derive that it is sufficient to show the following proof obligation:
\begin{align*}
\stateFormula^-& \land \preMonitor\left(\bar{x}^-,\bar{x}_1\right)
~\implies
\exists \bar{x}^+\Big(
    \controllerMonitor\left(\bar{x}_1,\bar{x}^+\right) \land\\
&\forall \bar{x} \left(
\bar{x}=\bar{x}^+ \implies
\left[\postProgram\right]
\left(\controllerMonitor\left(\bar{x}^-,\bar{x}\right)\land\stateFormula^-\right)
\right)
\Big)
\end{align*}
We can then easily derive that the obligation above is implied by
\begin{align*}
&\stateFormula^- \land \preMonitor\left(\bar{x}^-,\bar{x}_1\right)
~\implies\exists \bar{x}^+\Big(
\controllerMonitor\left(\bar{x}_1,\bar{x}^+\right) \land\\
&\forall \bar{x} \left(
\bar{x}=\bar{x}^+ \implies
\forall \bar{x} \left(
\postMonitor\left(\bar{x}^+,\bar{x}\right)
\implies
\left(\controllerMonitor\left(\bar{x}^-,\bar{x}\right)\land\stateFormula^-\right)
\right)
\right)
\Big).
\end{align*}
To this end, starting from latter formula, add $\left[\postProgram\right]$ before the inner $\forall \bar{x}$ via the vacuous axiom, then instantiate the inner $\forall \bar{x}$ with its outer version, apply $K$ and use that by construction
$\bar{x}=\bar{x}^+ \implies \left[\postProgram\right] \postMonitor\left(\bar{x}^+,\bar{x}\right)$.
Note that we can now remove $\stateFormula^-$ from the inner formula as it is implied by the left side of our implication.
Simplifying the formula further, renaming variables, and adding implicit all quantifiers yields the following formula, which implies our first proof obligation:
\begin{align*}
\forall \bar{x}_0~&\forall \bar{x}_1~\exists \bar{x}_2~\forall \bar{x}_3\enspace
\stateFormula\left(\bar{x}_0\right) \land \preMonitor\left(\bar{x},\bar{x}_1\right)
~\implies\\
&\left(
\controllerMonitor\left(\bar{x}_1,\bar{x}_2\right) \land
\left(
\postMonitor\left(\bar{x}_2,\bar{x}_3\right)
\implies
\controllerMonitor\left(\bar{x}_0,\bar{x}_3\right)
\right)
\right).
\end{align*}
This formula is guaranteed by our assumption that \Cref{fml:arithmetic_winning_strategy} is valid.

\paragraph*{(B)}
The second proof obligation reads:
\[
\stateFormula^- \land \controllerMonitor\left(x^-,x\right) \implies \left[\environmentProgram\right] \stateFormula.
\]
We note again that this formula is free of dualities and thus performs the proof in \dL{}.
To this end, consider any state $\mu \models \stateFormula^- \land \controllerMonitor\left(x^-,x\right)$.
Then, by construction of $\controllerMonitor$ we know there exists a state $\nu$ such that $\nu \models \stateFormula$ and $\left(\nu,\mu\right) \in \llbracket \controllerProgram\rrbracket$.
From \Cref{fml:safety_predicate} we already know that due to $\nu \models \stateFormula$ it must hold that for all $\left(\nu,\omega\right) \in \llbracket \controllerProgram;\environmentProgram\rrbracket$ we have $\omega \models \stateFormula$.
Thus, by definition of sequential composition for state transitions, we get that for all $\left(\mu,\hat{\omega}\right)\in\llbracket \environmentProgram\rrbracket$ it holds that $\hat{\omega} \models \stateFormula$.
Thus $\nu \models \left[\environmentProgram\right]\stateFormula$.
Consequently, this second proof obligation is valid due to our assumptions.
\end{proofE}

\paragraph*{Liveness}
When both $\preProgram$ and $\postProgram$ are set to the skip program $?\left(\top\right)$, the robustness property reduces to a standard \emph{liveness} property, i.e., whether the control envelope (without perturbation) contains a safe action in \emph{all} reachable states. In this sense, robustness can be seen as liveness under input/output perturbations.

\paragraph*{Robot Example}
\looseness=-1
As noted in \Cref{overview}, the robot restricted to moving toward the wall may crash if velocity computation is even slightly perturbed.
Using the formalization from \Cref{subsec:methods:example} and our robustness notion, we can state that
the $\envelopeExactOneDirOneWall$ is not robust w.r.t. the angelic perturbation $(\preProgramParam{1},\postProgramParam{1})$, as confirmed by a counterexample to the corresponding \dGL{} formula in \Cref{def:robust_control_envelopes} in KeYmaera~X
(\ifextension{see \Cref{apx:running-example}}{see also extended version's appendix}).
In contrast, for $\delta_v \leq V_{\text{max}}/2$, we prove via \Cref{thm:robust_envelope} that $\envelopeExactTwoDirOneWall$ is robust w.r.t. angelic perturbation $(\preProgramParam{1},\postProgramParam{1})$.
Similarly, $\envelopeExactTwoDirTwoWall$ is robust under the additional assumption that $\delta_v*T \leq W/2$, ensuring sufficient distance between the \emph{two} walls for safe operation under worst-case velocity perturbations.

\subsection{Implementation Safety under Perturbation}
\label{subsec:method:verification}
We have shown how to check the robustness of control envelopes by proving that each state admits \emph{one} safe action. However, this is insufficient for concrete implementations, whether idealized or fixed-point, where \emph{every} chosen action must ensure safety.  

In \dL{} terms, given a before-after predicate for an implementation $\implementationPredicateSym$ and its \dL{} representation $\alpha_{\text{refl}}\left(\implementationPredicateSym\right)$, safety under perturbation requires proving:
\begin{equation}
\label[formula]{fml:implementationSafety}
\precondition \implies \left[\left(
\preProgram;\alpha_{\text{refl}}\left(\implementationPredicateSym\right);\postProgram;\environmentProgram
\right)^*\right]\postcondition
\end{equation}

\looseness=-1
By translating angelic perturbations into real arithmetic~\cite{DBLP:journals/fmsd/MitschP16,DBLP:conf/ijcar/PrebetP24}, we derive a decidable, real arithmetic condition for safety under perturbation
\ifextension{(see proof on \cpageref{{proof:correctness_under_perturbation}})}{(see proof in extended version)}.
\begin{theoremE}[Safety under Perturbation][end,restate,text link=]
\label{thm:correctness_under_perturbation}
\looseness=-1
Let \controllerProgram{} be like in \Cref{def:robust_control_envelopes} and $\stateFormula$ be the inductive invariant used to prove $\safetyPredicateSym\left(\dots\right)$.
For angelic perturbation $\preProgram,\postProgram$ and monitor formulas $\controllerMonitor{},\preMonitor{},\postMonitor{}$ for $\controllerProgram{},\preProgram{},\postProgram{}$ the validity of the formula below implies \Cref{fml:implementationSafety}.
\begin{align}
\nonumber
\forall \bar{x}_0&~\forall{x}_1~\forall \bar{x}_2~\forall \bar{x}_3\enspace
\big(\stateFormula\left(\bar{x}_0\right) \land \preMonitor\left(\bar{x}_0,\bar{x}_1\right)
\land\\
& \implementationPredicate{\bar{x}_1}{\bar{x}_2} \land
 \postMonitor\left(\bar{x}_2,\bar{x}_3\right)\big)
\quad \implies\enspace \controllerMonitor\left(\bar{x}_0,\bar{x}_3\right)\label[formula]{fml:implementation_correct}
\end{align}
\end{theoremE}
\begin{proofE}
\label{proof:correctness_under_perturbation}
For our proof, we assume the validity of \Cref{fml:implementation_correct}.
We now begin a proof of \Cref{fml:implementationSafety} via the \dL{} proof calculus~\cite{Platzer2012,DBLP:journals/jar/Platzer17}.
First, we apply the loop invariant rule for \stateFormula{} to \Cref{fml:dgl_safety_abstract}.
The formulas $\precondition \implies \stateFormula$ and $\stateFormula \implies \postcondition$ immediately follow from our assumption that \Cref{fml:safety_predicate} was proven via the invariant \stateFormula{}.
From this assumption, we also know that:
\[
\stateFormula \implies \left[\controllerProgram;\environmentProgram\right] \stateFormula.
\]
It remains to show that:
\[
\stateFormula \implies \left[\preProgram;\alpha_{\text{refl}}\left(\implementationPredicateSym\right);\postProgram;\environmentProgram\right] \stateFormula.
\]
Let $V$ be the set of all variables appearing in the formula above.
For each variable $x\in V$ part of this formula, we now add a fresh, discrete ghost variable $x^-$, which is assigned to the same value as $x$.
Simple reasoning then yields that the above formula can be shown by showing the validity of the formula\footnotemark{}
\footnotetext{
Where $\bar{x}^- = \bar{x}\enspace\equiv\enspace\bigwedge_{x \in V} x^- = x$
}
\[
\stateFormula \land \stateFormula^- \land \bar{x}^- = \bar{x} \implies \left[\preProgram;\alpha_{\text{refl}}\left(\implementationPredicateSym\right);\postProgram;\environmentProgram\right] \stateFormula
\]
where $\stateFormula^-$ is the formula $\stateFormula$ where each apperance of $x \in V$ has been replaced by $x^-$.
Via the monotonicity rule, we can decompose this into two proof obligations from hereon referred to as (A) and (B):

\paragraph*{(A)}
The first proof obligation reads as follows:
\begin{align*}
\stateFormula & \land \stateFormula^- \land \bar{x}^- = \bar{x} \implies \\
&\left[\preProgram;\alpha_{\text{refl}}\left(\implementationPredicateSym\right);\postProgram\right]
\left(\stateFormula^- \land \controllerMonitor\left(\bar{x}^-,\bar{x}\right)\right).
\end{align*}
First, note that $\bar{x}^-$ is not part of the program inside the modality. Consequently, we can prove the preservation of $\stateFormula^-$ via $\left[\right]\land$ and the vacuous axiom, leaving the following proof obligation:
\begin{align*}
\stateFormula& \land \bar{x}^- = \bar{x} \implies \\
&\left[\preProgram;\alpha_{\text{refl}}\left(\implementationPredicateSym\right);\postProgram\right]
\left(\controllerMonitor\left(\bar{x}^-,\bar{x}\right)\right).
\end{align*}
Via the application of the $\left[;\right]$ rule, this can be transformed into the following formula (for clarity, we renamed $\bar{x}^-$ to $\bar{x}_0$):
\begin{align*}
\stateFormula &\left(\bar{x}_0\right) \land \bar{x}_0 = \bar{x} \implies \\
&\left[\preProgram\right]\left[\alpha_{\text{refl}}\left(\implementationPredicateSym\right)\right]\left[\postProgram\right]
\left(\controllerMonitor\left(\bar{x}_0,\bar{x}\right)\right).
\end{align*}
As previously seen in the proof of \Cref{thm:robust_envelope}, we can now use a monotonicity argument and the insight that by construction $\bar{x}_0 = \bar{x} \implies \left[\preProgram\right]\preMonitor\left(\bar{x}_0,\bar{x}\right)$ to reduce the proof obligation to 
\begin{align*}
\stateFormula\left(\bar{x}_0\right) & \land \preMonitor\left(\bar{x}_0,\bar{x}\right) \implies \\
&\left[\alpha_{\text{refl}}\left(\implementationPredicateSym\right)\right]\left[\postProgram\right]
\left(\controllerMonitor\left(\bar{x}_0,\bar{x}\right)\right).
\end{align*}
By reintroducing a set of fresh ghost variables $\bar{x}_1$ such that $\bar{x}_1=\bar{x}$ we can derive $\preMonitor\left(\bar{x}_0,\bar{x}_1\right)$ and subsequently replace $\alpha_{\text{refl}}\left(\implementationPredicateSym\right)$ via a Monotonicity argument for $\implementationPredicate{\bar{x}_1}{\bar{x}}$.
A similar Ghost+Monotonicity argument for $\postProgram$ then yields that we can prove the original proof obligation of (A) by showing the validity of the following formula:
\begin{align*}
\stateFormula\left(\bar{x}_0\right) &\land
\preMonitor\left(\bar{x}_0,\bar{x}_1\right)  \land
\implementationPredicate{\bar{x}_1}{\bar{x}_2} \land
\postMonitor\left(\bar{x}_2,\bar{x}_3\right)\\
&
\implies 
\left(\controllerMonitor\left(\bar{x}_0,\bar{x}_3\right)\right).
\end{align*}
This corresponds to our assumption from \Cref{fml:implementation_correct}.

\paragraph*{(B)}
The second proof obligation reads:
\begin{align*}
\stateFormula^- \land \controllerMonitor\left(x^-,x\right) \implies\left[\environmentProgram\right] \stateFormula.
\end{align*}
To this end, we refer the reader to part (B) of the proof for \Cref{thm:robust_envelope}, which proves the same statement under the same assumptions.
\end{proofE}
Phrased differently, proving \Cref{fml:implementation_correct} shows that \implementationPredicateSym{} encodes a \emph{winning strategy} for demon in the game of \Cref{fml:dgl_safety_abstract} (with \controllerProgram{} substituted by $\alpha_{\text{refl}}\left(\implementationPredicateSym\right)$).
For NNs, verification is typically done using NN verification tools, not general real arithmetic solvers. In this setting, \Cref{fml:implementation_correct} (with $\implementationPredicateSym$ omitted) serves directly as the NN's specification and can be simplified further %
(\ifextension{see \Cref{apx:simplified_conditions}}{see extended version}).

\paragraph*{Robot Example}
\looseness=-1
We previously showed that $\envelopeExactOneDirOneWall$ and $\envelopeExactTwoDirOneWall$ are robust under angelic perturbation $(\preProgramParam{1},\postProgramParam{1})$.
We now examine whether the idealized implementation $\implementationPredicateRobotExact{p}{v^+}$ (\Cref{subsec:methods:example}) is safe under perturbation.

To this end, we instantiate \Cref{fml:implementation_correct} for
all combinations of $\left\{\envelopeExactTwoDirOneWall,\envelopeExactTwoDirTwoWall\right\} \times \left\{(\preProgramParam{1},\postProgramParam{1}),(\preProgramParam{2},\postProgramParam{2})\right\}$ w.r.t. $\implementationPredicateRobotExactSym{}$, yielding
four real arithmetic formulas. Each formula's validity implies the safety of implementation $\implementationPredicateRobotExactSym{}$ w.r.t. its environment (one/two walls) and chosen perturbation.

For $T=1.0,V_{\text{max}}=10,W=100,M=9.5$ and $\delta_p=\delta_v=0.25$ all 4 formulas are proven using KeYmaera~X.
Hence, any fixed-point implementation differing from $\implementationPredicateRobotExactSym{}$ by at most 0.25  and with sensor perturbations of $\leq 0.25$ remains safe within the fixed domain $p \in \left[0,100\right]$.
We now use the
safety result in \Cref{subsec:method:implementation} to generate safe, efficient fixed-point code.

Importantly,
we previously disproved the robustness of $\envelopeExactOneDirOneWall$ w.r.t. $(\preProgramParam{1},\postProgramParam{1})$.
Hence, as robustness is a necessary condition, no matter the parameter choice, we cannot prove safety of $\implementationPredicateRobotExactSym{}$ w.r.t. $\envelopeExactOneDirOneWall$ under the given angelic pertrubations.
Without checking robustness,
a failed safety proof under perturbation leaves unclear if this is due to a limitation of the fixed-point implementation or of the chosen \dL{} model.

\subsection{Implementation Synthesis}
\label{subsec:method:implementation}
With $\implementationPredicateRobotExact{p}{v^+}$ proven safe under bounded perturbations, we use state-of-the-art mixed-precision tuning to synthesize a fixed-point implementation that stays within the chosen bound $\delta_v$. 
For this, we use Daisy~\cite{Daisy} (also supports uniform precision) that performs tuning using interval or affine arithmetic. It generates C++ code with mixed-precision fixed-point arithmetic (up to 64-bit width), which can be directly compiled by Xilinx~\cite{websiteOfVivado} to measure runtime in machine cycles.

However, Daisy does not natively support vectors, matrices, or loops, common in NNs. It requires unrolled structures and loops and also generates fully unrolled code, which may overwhelm Xilinx~\cite{Aster}. To address this, we add a pre-processing step to automatically unroll standard NN structures for Daisy and a post-processing step to reintroduce the structures and loops after tuning. This enables Daisy to handle our neural networks effectively. 
Note that we also considered the tool Aster~\cite{Aster}, which supports NN-specific code.
But it is limited to 32-bit precision due to solver constraints and becomes infeasible for deeper networks due to imprecise range overapproximations, making it unsuitable for our purpose.

For our robot example, we first used extended Daisy to generate a 32-bit uniform fixed-point implementation of $\implementationPredicateRobotExact{p}{v^+}$, yielding a worst-case roundoff error of $5.37 \times 10^{-6}$ and a latency of 68 cycles. Since we proved safety under perturbations up to $\delta_v = 0.25$, Daisy can synthesize a more efficient mixed-precision implementation using at most 17 bits, reducing latency to 30 cycles while still ensuring safety.

\subsection{Limitations}
As stated in \Cref{thm:robust_envelope}, 
\Cref{fml:arithmetic_winning_strategy}
implies controller robustness.
However, if \Cref{fml:arithmetic_winning_strategy} is not valid, this does not necessarily mean the controller is non-robust.
A counterexample $\bar{x}_0$ satisfying $\precondition$ is a concrete violation, but one outside of $\precondition$ may be spurious; robustness might still hold under a \emph{stronger} loop invariant that excludes $\bar{x}_0$.

Thus, our control envelope approach is, in general, incomplete.
However, many \dL{} proofs provide
guarantees for the weakest assumptions and thus set the invariant region as the initial condition proving safety results for very permissive control envelopes: If $\precondition \iff \stateFormula$ (and not only $\precondition \implies \stateFormula$), then any counterexample is a concrete violation w.r.t. actions deemed safe by $\controllerProgram{}$.

Finally, our method relies on proven, safe \dL{} control envelopes.
Fortunately, a wide range of envelopes exists across applications~\cite{DBLP:journals/sttt/JeanninGKSGMP17,DBLP:journals/tcad/KabraMP22,DBLP:conf/icfem/PlatzerQ09,DBLP:journals/tiv/SelvarajAF23,DBLP:conf/fmics/WuRF20,DBLP:journals/ral/BohrerTMSP19,DBLP:conf/asm/PrebetTP25}, and recent work~\cite{DBLP:conf/tacas/KabraLMP24} shows that their synthesis can be automated.

\section{Evaluation}\label{sec:evaluation}
We evaluate our methodology on three realistic case studies to demonstrate its practical applicability. We begin with Adaptive Cruise Control (ACC), verifying safety for both discrete and continuous action spaces. 
We then apply our approach to the more complex and safety-critical Vertical Airborne Collision Avoidance System (VCAS), specifically focusing on two neural networks-Do-Not-Climb (DNC) and Do-Not-Descend (DND)—which issue collision avoidance advisories in case the current advisory is DNC/DND.

While the NNs for continuous ACC and VCAS are taken from prior work~\cite{Teuber2024,JulianKDVCASSafe,julian2016policy}, the discrete ACC networks were trained specifically for this paper.
The continuous ACC network has 256 ReLU nodes and over 12k parameters; the discrete variant has the same architecture but has 3 output nodes instead of 1.
Each of the two networks used in the VCAS case study contains 6x45 ReLU nodes.
\Cref{tab:eval_results_overview} summarizes our experimental results.

\subsection{Experimental Setup}
\looseness=-1
All experiments were conducted on an Ubuntu 22.04 system with an AMD Ryzen 7 CPU and 32 GB RAM. Safety proofs in differential dynamic logic and real arithmetic were performed interactively using KeYmaeraX\cite{Fulton2015}, which internally relies on Mathematica~\cite{Mathematica} for quantifier elimination.
NN verification was done using $N^3V$\cite{Teuber2024,samuel_teuber_2024_13922169}, which supports polynomial specifications with arbitrary propositional structure. Internally, $N^3V$ is based on 
Z3%
~\cite{DBLP:conf/tacas/MouraB08,DBLP:conf/cav/BjornerN24},
PicoSAT%
~\cite{PicoSAT} and the linear specification NN verifier 
nnenum~\cite{bakOverapprox}.
For mixed-precision code generation and FPGA synthesis, we used Daisy's March 2, 2021 version (with no major updates since then)
and Xilinx’s Vitis HLS~\cite{websiteOfVivado} (version 2023.1), downloaded on September 24, 2024.

\begin{table}[t]
    \centering
    \caption{Summary of Results. $\delta$: worst-case perturbation/error bound; Time~($\mathbb{R}$): verification time with reals; Time~($\delta$): verification time under perturbation; NN Synthesis Time: time to generate an NN with error bound $\delta$; Cycles: latency reported by Xilinx.}
    \begin{tabular}{l|r||r|r||r|r}
         \multirow{2}{*}{\textbf{Case Study}} & \multirow{2}{*}{$\mathbf{\delta}$} & \multicolumn{2}{c||}{\textbf{NN Verification}} &
         \multicolumn{2}{c}{\textbf{NN Synthesis}} \\\cline{3-6}
         &&\textbf{Time ($\mathbb{R}$)} & \textbf{Time ($\delta$)} &
         \textbf{Time} & \textbf{\# Cycles}\\\hline\hline
         \textbf{ACC (cont.)} & 1.0\hphantom{0} & %
          2.02m & 9.3m & 1.53h & 567\\
         \textbf{ACC (disc.)} & 0.01 & 59s & 1m & 2.01h & 579\\
         \textbf{VCAS} DNC & $25^{-3}$ & %
         18.92m& 27.45m & 1.73h & 655\\
         \hspace{7.5mm} DND & $25^{-3}$ & %
         15.75m & 19.1m
         & 1.82h & 656
    \end{tabular}
    \label{tab:eval_results_overview}
\end{table}

\subsection{Case Study 1: Continuous Adaptive Cruise Control}
For continuous ACC, prior work~\cite{Teuber2024} provides a \dL{} safety guarantee for NN verification.
In this setting, an \emph{ego car} (following a \emph{front car}) chooses a relative acceleration $a_{\text{rel}}$ based on current relative position $p_{\text{rel}}$ and velocity $v_{\text{rel}}$.
The safety goal is to prevent collisions while ensuring the ego car does not fall behind.
Previous work had also established a real-valued safety guarantee for a validated NN~\cite[Sec. 5]{Teuber2024}, raising the question whether the same safety guarantee holds in a finite precision setting.
To this end, we formalize an angelic perturbation as $\left(?\left(\top\right),\postProgram^{\text{acc}}\right)$ with:
\begin{equation*}
\postProgram^{\text{acc}}~\equiv~\left(\varepsilon_{a_{\text{rel}}}\coloneqq *;?\left(\left|\varepsilon_{a_{\text{rel}}}\right| \leq \delta_{a_{\text{rel}}}\right); a_{\text{rel}} \coloneqq a_{\text{rel}}+\varepsilon_{a_{\text{rel}}}\right)
\end{equation*}
\paragraph*{Envelope Robustness}
Initial analysis showed that the control envelope derived before is not robust under $\left(?\left(\top\right),\postProgram^{\text{acc}}\right)$.
It assumes actions in the range $a_{\text{rel}} \in \left[-B,A\right]$, where $-B$ is the maximum braking acceleration, and guarantees safety as long as braking with $-B$ avoids a crash. However, this relies on the \emph{precise} execution of $-B$, which is unrealistic under finite precision. We therefore revised the control envelope to reflect this limitation and formally proved its safety and robustness in KeYmaera~X.

\paragraph*{NN Verification}
Using the revised control envelope, we derived the specification for the NN as described in \Cref{subsec:method:implementation} and verified the NN using $N^3V$.
\Cref{tab:eval_results_overview} reports 
the verification times for both the original (Time~$\mathbb{R}$) and perturbation-aware (Time~$\delta$) specifications.
While verification under perturbation takes longer, it remains feasible. 
The longer time is likely due to the increased propositional complexity of the specification, leading to more queries to the underlying verifier nnenum.

\paragraph*{Code Generation}
Finally, we used Daisy~\cite{Daisy} to generate an efficient mixed-precision fixed-point C++ implementation of the NN that meets the error bound $\delta = 1.0$. To compute latency, we compiled the generated code with Xilinx Vivado HLS~\cite{websiteOfVivado} for a standard FPGA architecture. Note that this latency is exact, i.e., we do not consider noisy runtime measurements from actual hardware.

With our pre- and post-processing pipeline, Daisy successfully generated code with an error of 1.0. 
While uniform precision required 32 bits, mixed-precision tuning produced an implementation using a mix of 31- and 32-bit variables, reducing total bit usage by 3.66\% compared to the uniform version. The latency remained at 567 cycles, as the limited reduction in bit widths was not sufficient to impact the overall execution time.  We note that retaining the fully unrolled version could potentially reduce latency, but the size may overwhelm Xilinx.

\paragraph*{Discussion}
The continuous ACC case study illustrates that the proposed approach presented in this paper applies effectively to realistic regression-based NN control systems, handling roundoff errors. %

While NN verification typically scales with the number of $\mathrm{ReLU}$ nodes, mixed precision tuning and code generation scale with the number of parameters of the NN, \emph{including} all weights and biases.
Since both steps are performed only once, our method remains practical even for networks with 256 $\mathrm{ReLU}$ nodes and over 12K parameters.

Most importantly, the proposed approach enhances \emph{warranted} trust in NN controllers by ensuring \emph{infinite-time horizon} safety and bridging the gap between theoretical guarantees in real arithmetic and their efficient, deployable code on FPGAs.

\subsection{Case Study 2: Discrete Adaptive Cruise Control}
In this case study, the NNs are classifiers selecting one of three actions ($a_{\text{rel}} \in {-B, 0, A}$) via an argmax over three outputs.
Finite-precision analysis is more challenging here, as \emph{any} error can flip decisions for \emph{some} inputs near classification boundaries. For instance, if $a_{\text{rel}} = -B$ is selected for inputs $x$ such that $\langle w,y\rangle \leq b$, then for input values with $\langle w, y\rangle = b$, even infinitesimal perturbations can flip the action.

Hence, in this case study, the key question is not \emph{if} the NN's behavior changes under perturbations but \emph{where} it changes.
For the chosen control envelope and idealized implementation we verified safety under perturbation for a trained NN for $\delta=0.01$ which simultaneously shows the envelope's robustness.

With a perturbation bound of 0.01, Daisy estimated that 32-bit uniform precision was sufficient to meet the error requirements.  We then used Daisy to synthesize a mixed-precision implementation, assigning 29 to 32 bits across variables. The mixed-precision tuning took 2.01 hours and produced code with a latency of 579 cycles, achieving a 0. 69\% reduction in latency and a 2.14\% improvement in total bit usage compared to the 32-bit uniform precision implementation.
\subsection{Case Study 3: Vertical Airborne Collision Avoidance}
Our third case study focuses on the Vertical Airborne Collision Avoidance System (VCAS), developed through multiple Federal Aviation Administration (FAA)-supervised research efforts~\cite{holland2013optimizing,kochenderfer2012next}.
It aims to prevent Near Mid-Air Collisions (NMACs), defined as aircraft coming within 500 feet horizontally or 100 feet vertically, by identifying dangerous trajectories and issuing timely advisories to adjust flight paths.

Given that hazardous situations often develop rapidly, the advisory system must provide \emph{correct} decisions. 
Previous work has formalized this CPS in \dL{}~\cite{DBLP:journals/sttt/JeanninGKSGMP17} and proposed NN controllers for use~\cite{julian2016policy,JulianKDVCASSafe}. Although the full combination of these NNs is unsafe, two specific NNs (DNC and DND) were proven safe under idealized real-valued semantics~\cite{Teuber2024}. Like in the discrete ACC, these controllers work in a discrete action space with similar challenges under finite precision.

VCAS represents a critical application where providing formal safety guarantees for fixed-point implementations---not just idealized models---is crucial. To this end, we applied our methodology to the two verified NNs, constructing an angelic perturbation model analogous to $\postProgram^{\text{acc}}$ to reflect realistic numerical and actuation uncertainties. We then derived the corresponding NN safety specifications.

Using $N^3V$, we verified both NNs under this new specification in comparable time to the original setting. Daisy's analysis showed that 42-bit uniform precision is required to meet the perturbation bound of $25^{-3}$. We then synthesized mixed-precision implementations in 1.73 hours (DNC) and 1.82 hours (DND). While the latencies remained the same at 655 and 656 cycles w.r.t. the uniform baselines, mixed-precision versions have bit savings of 6.18\% and 5.2\%, respectively.

\subsection{Discussion}
Our three case studies demonstrate that our approach scales to safety-critical, real-world systems like VCAS and applies to both regression and classification-based NN Control Systems.
Importantly, our methodology provides \emph{global} safety guarantees for the NN, which are generally more challenging to obtain and scale less than local robustness guarantees. 
However, many control applications of interest rely on NNs of comparable size~\cite{ARCH21:ARCH_COMP21_Category_Report_Artificial}, making our approach practically relevant.

\section{Related Work}
\label{sec:related}
\paragraph*{Safety Analysis}
While many techniques guarantee safety for Cyber-Physical Systems (CPS) control, most of them omit finite-precision or sensor/actuation errors. In principle, \dL{}\cite{DBLP:journals/jar/Platzer08,Platzer2012,DBLP:journals/jar/Platzer17} supports reasoning about such errors; however, case studies often avoid it due to complexity~\cite{DBLP:journals/sttt/JeanninGKSGMP17,DBLP:conf/fmics/WuRF20,DBLP:journals/tiv/SelvarajAF23}. Our approach addresses this gap by enabling robustness analysis of control envelopes from such studies. 

VerSAILLE~\cite{Teuber2024} is the only method using \dL{} guarantees for NNCSs.
This enables \emph{infinite-time} horizon guarantees.
In contrast, closed-loop reachability~\cite{Forets2019JuliaReachReachability,Schilling2021,tran2019star,tran2020neural,9093970,IvanovACM21,Huang2019,Fan2020,Dutta2019,Sidrane2021,Akintunde2022,Polar2022} ensures safety only over a finite horizon, limiting inductive reasoning~\cite[Appendix E.2]{Teuber2024}. Similarly, barrier certificates~\cite{noroozi2008generation,DBLP:conf/corl/RichardsB018,DBLP:conf/tacas/PeruffoAA21,DBLP:journals/csysl/AbateAGP21,DBLP:conf/hybrid/AbateAEGP21} lack the exact reasoning supported by VerSAILLE, and typically cover smaller regions.
Unlike our work, all mentioned NNCS verification works \emph{ignore} the effects of fixed-point arithmetic which are essential for \emph{realistic} safety guarantees.
Unlike~\cite{harikishan2024towards}, which integrates closed-loop reachability with bounding fixed-point errors, we extend VerSAILLE and can thus provide stronger, infinite-time horizon guarantees.

Recent work~\cite{HabeebConformanceClosedLoop} analyzes divergence between closed-loop NN controllers using $\varepsilon$-equivalence~\cite{DBLP:conf/cp/BuningKS20,paulsen_reludiff_2020} and Lipschitz analysis, but focuses on finite-horizon settings and does not address sound quantization. For a broader comparison between closed-loop verification, barrier certificates, and VerSAILLE, we refer to~\cite[Sec. 5, 6, Appendix E.2]{Teuber2024}.

Efforts to verify quantized NNs~\cite{henzinger2021scalable,DBLP:conf/kbse/ZhangZCSZCS22,DBLP:conf/aaai/000200DWZB24} use bit-vector constraints.
These tools are incompatible with the real-valued polynomial constraints required by VerSAILLE, limiting their use for infinite-horizon NNCS verification.

\paragraph*{Mixed Precision Tuning}
For sound mixed-precision tuning, Daisy~\cite{Daisy} (used in this work) and POPiX~\cite{bessai2022fixed} support fixed-point arithmetic, while FPTuner~\cite{Chiang2017}, Salsa~\cite{damouche2018mixed}, and POP~\cite{adje2021fast} focus on floating-point programs. Unlike Daisy, POPiX frames tuning as an ILP problem but relies on dynamic range analysis, which compromises soundness. Daisy also integrates expression rewriting to improve mixed-precision tuning, though this added step increases computational cost, making it harder to scale for larger benchmarks.

\paragraph*{Neural Network Quantization}
State-of-the-art NN quantization~\cite{KumarS020,ParkKY18,SharmaPSLCCE18,SongFWJ0JL20} typically target NN classifiers, optimizing accuracy without formal guarantees. Recent efforts, such as Aster~\cite{Aster}, automate mixed-precision tuning for NN controllers under error bounds but suffer from imprecise integer range estimation, limiting their usability for \emph{deeper} networks. Popinns~\cite{khalifa2024rigorous} also synthesizes bounded fixed-point implementations but relies on dynamic analysis and assumes access to a floating-point model, unlike our approach, where the goal is sound efficient code generation.

\paragraph*{Combined Approaches}

VeriPhy~\cite{DBLP:conf/pldi/BohrerTMMP18} synthesizes sandboxes based on \dL{} contracts and interval-based fallback controllers. However, it provides guarantees around, not \emph{of}, the controller itself. Extending VeriPhy to realistic NNs would require substantial changes, as NNs are not amenable to interactive proofs. We view VeriPhy as orthogonal to our approach, applicable when direct NN verification is impossible.

Finally, for code generation, we use Xilinx Vivado~\cite{websiteOfVivado} for traditional and NN controllers. For neural controllers, tools like FloPoCo~\cite{de2019reflections} can generate optimized FPGA-based dot products, which we leave as future work.

\section{Conclusion}
This paper addresses the critical gap between theoretical infinite-horizon safety guarantees for NN-controlled systems and their practical finite-precision implementations. Since real-valued guarantees may fail under finite precision, we formulated control envelope robustness as a game between a \emph{good} Demon and a \emph{bad} Angel. Verifying infinite-horizon safety thus reduces to showing that the NN implements a winning strategy while keeping fixed-point implementation errors within allowable perturbations. Our approach relies on decidable real-arithmetic conditions, avoiding the need for undecidable \dGL{} or \dL{} validity proofs.

Our experiments on realistic NN control systems, from automotive to aeronautics, covering both regression- and classification-based NN architectures, demonstrate that accounting for perturbations introduces no significant overhead compared to perturbation-free verification, though costs may increase for more complex systems. While this paper focuses on NN controllers, the proposed technique naturally extends to other controller types.
Overall, our work represents 
a significant step toward fostering \emph{warranted} trust in real-world NN control systems.

\bibliographystyle{ieeetr}
\bibliography{main.bib}
\ifextension{
\appendix
\section*{\dL{} Model of the Running Example}
\label{apx:running-example}
The robot's physical behavior, or \emph{plant model},
can be modelled as follows where the evolution constraint $t \leq T$ ensures the controller takes actions at least every $T$ seconds:
\[
\environmentProgram~\equiv~t \coloneqq 0\; \left(p'=-v,t'=1\&t\leq T\right).
\]
The following (real-valued) \dL{} safety results have been derived in KeYmaera~X with the definitions from \Cref{fig:controller_safety}:

\begin{lemma}
The following \dL{} formulas are valid:
\begin{align*}
\preExactOneWall \rightarrow
\left[\left(
    \envelopeExactOneDirOneWall;
    \environmentProgram
\right)^*
\right]\postExactOneWall\\
\preExactOneWall \rightarrow
\left[\left(
    \envelopeExactTwoDirOneWall;
    \environmentProgram
\right)^*
\right]\postExactOneWall\\
\preExactTwoWall \rightarrow
\left[\left(
    \envelopeExactTwoDirTwoWall;
    \environmentProgram
\right)^*
\right]\postExactTwoWall
\end{align*}
\end{lemma}
\begin{proof}
    Proofs performed in KeYmaera~X.
\end{proof}

\paragraph*{Intuition for non-robustness}
In the paper's main text, we note that $\envelopeExactOneDirOneWall$ is not robust w.r.t. $\left(\preProgramParam{1},\postProgramParam{1}\right)$.
To this end, consider the case where $p=0$:
In this case, the robot must not, under any circumstances, move forward as this would push the robot directly into the obstacle.
Since the envelope $\envelopeExactOneDirOneWall$ only allows $v \geq 0$, this means the robot must choose $v=0$.
However, $v$ is subsequently perturbed by $\postProgramParam{1}$ which can lead to a velocity $v=\delta_v>0$.
This would crash the robot into the obstacle.
Consequently, in this state, the robot does not have \emph{any} action inside the envelope that would keep it safe.
We also formalized this in KeYmaera X by disproving the formula from \Cref{def:robust_control_envelopes} for the given instantiation.

\section*{Simplified Robustness \& Safety Conditions}
\label{apx:simplified_conditions}

\subsection*{Robustness under Bounded Perturbations}
A typical case is one where both inputs and outputs of the controller are perturbed by some $\left|\varepsilon\right|\leq \delta$.
Importantly, this formulation can capture all three influence factors (I1) through (I3) of imprecision.
For example, for finite precision computation, we bound the error of the computed outcome with $\delta$.
For the angelic perturbation $\left(\preProgramParam{2},\postProgramParam{2}\right)$ from \Cref{subsec:method:perturbation}, we can reduce the criterion in \Cref{fml:arithmetic_winning_strategy} to a simplified format that also generalizes to the multidimensional setting:
\begin{align*}
\forall p~&\forall\varepsilon_p~\exists v^+~\forall \varepsilon_v\enspace
\stateExactTwoDirTwoWall\left(p\right) \land
\left|\varepsilon_p\right|\leq\delta_p
\implies\\
&\quad\monitorExactTwoDirTwoWall\left(p+\varepsilon_p,v^+\right)\land
\left(
\left|\varepsilon_v\right|\leq\delta_v
\implies
\monitorExactTwoDirTwoWall\left(p,v+\varepsilon_v\right)
\right).
\end{align*}
where $\monitorExactTwoDirTwoWall$ is the controller monitor for $\envelopeExactTwoDirTwoWall$ and $\stateExactTwoDirTwoWall\equiv\preExactTwoWall$ is the inductive invariant for the safety proof of $\stateExactTwoDirTwoWall$
The validity of the above formula then implies \Cref{fml:arithmetic_winning_strategy}.
In essence, disproving the validity of this formula amounts to finding concrete perturbation values $\varepsilon_p,\varepsilon_v$ that disprove safety.

\subsection*{Safety under Bounded Perturbations}
Again assumuing the angelic perturbation $\left(\preProgramParam{2},\postProgramParam{2}\right)$ from \Cref{subsec:method:perturbation}, we can reduce the property encoded in \Cref{fml:implementation_correct} to the following which reduces the number of quantified variables:
\begin{align*}
    &\forall p~\forall\varepsilon_p~\forall v^+~\forall \varepsilon_v\enspace
    \stateExactTwoDirTwoWall\left(p\right) \land\\
    &\left|\varepsilon_p\right|\leq\delta_p
    \land
    \monitorExactTwoDirTwoWall\left(p+\varepsilon_p,v^+\right)\land
    \left|\varepsilon_v\right|\leq\delta_v
    \implies
    \monitorExactTwoDirTwoWall\left(p,v+\varepsilon_v\right).
\end{align*}
This once again simplifies the analysis in practice.

\section*{Proofs}
\printProofs
}{}
\end{document}